\documentclass[aps,twocolumn,preprintnumbers,eqsecnum,prb]{revtex4}
\usepackage[dvips]{graphicx}
\usepackage[T1]{fontenc}
\usepackage{amsmath,amsbsy,amssymb,graphics}
\let\mathit=\mathcal

\begin{document}

\preprint{Accepted for publication in Physical Review B}

\title{Microscopic Theory of Skyrmions in Quantum Hall Ferromagnets}
\author{G. Tsitsishvili$^{1,2}$ and Z.F. Ezawa$^{1}$}
\affiliation{{}$^1$Department of Physics, Tohoku University, Sendai, 980-8578 Japan \\
{}$^2$Department of Theoretical Physics, A. Razmadze Mathematical Institute,
Tbilisi, 380093 Georgia}
\date{\today}

\begin{abstract}
We present a microscopic theory of skyrmions in the monolayer quantum Hall
ferromagnet. It is a peculiar feature of the system that the number density
and the spin density are entangled intrinsically as dictated by the W$%
_{\infty }$ algebra. The skyrmion and antiskyrmion states are constructed as
W$_{\infty }$-rotated states of the hole-excited and electron-excited
states, respectively. They are spin textures accompanied with density
modulation that decreases the Coulomb energy. We calculate their excitation
energy as a function of the Zeeman gap and compared the result with
experimental data.
\end{abstract}

\maketitle

\section{Introduction}

Quantum coherence in the quantum Hall (QH) system has proved to be a
fascinating subject during the past decade\cite{BookEzawa,BookDasSarma}.
Electron spins are spontaneously polarized even in the absence of the Zeeman
effect due to the exchange Coulomb interaction, leading to the QH
ferromagnet. A prominent characteristic is that a charged excitation is a
spin texture called skyrmion\cite{Sondhi93B}. It is a coherent excitation of
spins, and its excitation can be confirmed if the number of flipped spins is
found to be more than that of a hole excitation\cite{Barrett95L}. A
conventional way is to measure the increase of the activation energy by
tilting samples\cite{Schmeller95L,Aifer96L}, which is roughly proportional
to the number of flipped spins due to the Zeeman effect.

The skyrmion has been known\cite{BookRajaraman} to be a classical solution
to the nonlinear sigma model. Indeed, the concept of skyrmion was introduced%
\cite{Sondhi93B} into QH ferromagnets first in this context. Subsequently a
microscopic skyrmion state was considered to carry out a Hartree-Fock
approximation\cite{Fertig94B,MacDonald96B,Abolfath97B,Fertig97B}. In this
paper we elaborate this idea and present a microscopic theory of skyrmions
at the filling factor $\nu =1$, employing the framework of noncommutative
geometry\cite{EzawaX03B}. We also compare the result with the experimental
data\cite{Schmeller95L}.

The property of electrons becomes very peculiar when they are confined
within the lowest Landau level\cite{Girvin86B}. The electron position is
described solely by the guiding center $\mathbf{X}=(X,Y)$ subject to the
noncommutative relation, $[X,Y]=-i\ell _{B}^{2}$, with $\ell _{B}$ the
magnetic length. This noncommutativity is related with the so-called W$%
_{\infty }$ algebra\cite{Iso92PLB,Cappelli93NPB}. When the spin degree of
freedom is introduced, the algebraic property becomes the SU(2) extension of
the W$_{\infty }$ algebra\cite{Ezawa97B}, which we have named the W$_{\infty
}$(2) algebra\cite{EzawaX03B}. It implies an intrinsic entanglement of the
number density and the spin density of electrons, which amounts to a spin
excitation accompanied with a density modulation.

This paper is organized as follows. In Section \ref{SecDensiOpera} we
formulate the number density and the spin density of electrons confined
within the lowest Landau level. They form the W$_{\infty }$(2) algebra. In
Section \ref{SecHamil} we address the question whether there are states
having the same electron number but a lower excitation energy compared with
the hole state. In Section \ref{SecMicroSkyrm} we propose the skyrmion and
antiskyrmion states as W$_{\infty }$($2$)-rotated states of the hole-excited
and electron-excited states, respectively. They agree with the anzats\cite%
{Fertig94B} made in a Hartree-Fock approximation. Since a W$_{\infty }$($2$)
rotation modulates not only the spin density but also the number density, it
decreases the Coulomb energy of the excitation. In Section \ref{SecWaveFunct}
we derive the wave function of the skyrmion state. In particular we examine
the case where the wave function is factorizable in the electron
coordinates. We call such a skyrmion the factorizable skyrmion. In Section %
\ref{SecHardCore} we study the system governed by the hard-core interaction
instead of the Coulomb interaction, where the factorizable skyrmion is an
eigenstate of the Hamiltonian. However, it is shown that skyrmion
excitations are energetically unfavored once the Zeeman effect is taken into
account. In Section \ref{SecInterCoulo} we study the realistic system with
the Coulomb interaction together with the Zeeman interaction. The
factorizable skyrmion cannot be physical since its Zeeman energy diverges.
We propose a trial skyrmion which interpolates the hole and the factorizable
skyrmion. By minimizing the total energy we determine the state as well as
the excitation energy as a function of the Zeeman gap. The result is
compared successfully with the experimental data\cite{Schmeller95L} in
Section \ref{SecExperEvide}. In Section IX we re-examine the semiclassical
approximation\cite{Ezawa99L} in the present context of the microscopic
theory. It is shown that, though strictly speaking the factorizable skyrmion
cannot be physical, it still presents a reasonable approximation with an
appropriate cutoff of the divergent Zeeman energy. Section \ref{SecDiscu} is
devoted for the conclusion.

\section{Density Operators}

\label{SecDensiOpera}

Electrons in a plane perform cyclotron motion under strong magnetic field $%
B_{\perp }$ and create Landau levels. The coordinate $\mathbf{x}=(x,y)$ of
the electron is decomposed into the guiding center $\mathbf{X}=(X,Y)$ and
the relative coordinate $\mathbf{R}=(R_{x},R_{y})$, $\mathbf{x}=\mathbf{X}+%
\mathbf{R}$, where $R_{x}=-P_{y}/eB_{\perp }$ and $R_{y}=P_{x}/eB_{\perp }$
with $\mathbf{P}=(P_{x},P_{y})$ the covariant momentum. In the symmetric
gauge the decomposition implies%
\begin{equation}
X={\frac{1}{2}}x-i\ell _{B}^{2}{\frac{\partial }{\partial y}},\quad Y={\frac{%
1}{2}}y+i\ell _{B}^{2}{\frac{\partial }{\partial x}},  \label{RepreDiffe}
\end{equation}%
where $\ell _{B}=\sqrt{\hbar /eB_{\perp }}$ is the magnetic length. When the
cyclotron energy gap is large, thermal excitations across Landau levels are
practically suppressed at sufficiently low temperature. In such a system the
electron position is specified solely by the guiding center subject to the
noncommutative relation, 
\begin{equation}
\lbrack X,Y]=-i\ell _{B}^{2}.  \label{AlgebNC}
\end{equation}%
Due to this noncommutative relation an electron cannot be localized and
occupies an area $2\pi \ell _{B}^{2}$ called the Landau site. There are $%
N_{\Phi }=S/2\pi \ell _{B}^{2}$ Landau sites per one Landau level, where $S$
is the area of the system. It is equal to the number $N_{\Phi }=B_{\perp
}S/\Phi _{\text{D}}$ of flux quanta passing through the system, where $\Phi
_{\text{D}}=2\pi \hbar /e$ is the Dirac flux quantum. One Landau site may
accommodate two electrons with up and down spins according to the Pauli
exclusion principle. The filling factor is $\nu =N/N_{\Phi }=2\pi \ell
_{B}^{2}\rho _{0}$, where $N$ is the total number of electrons and $\rho
_{0}=N/S$ is the electron density. The system becomes incompressible,
leading to the integer QH effect, when the filling factor $\nu $ takes an
integer value.

In this paper we study the QH system at $\nu =1$. Thus all electrons are
assumed to be confined within the lowest Landau level. We define operators 
\begin{equation}
b=\frac{1}{\sqrt{2}\ell _{B}}(X-iY),\quad b^{\dag }=\frac{1}{\sqrt{2}\ell
_{B}}(X+iY)  \label{WeylXb}
\end{equation}%
with (\ref{RepreDiffe}). They obey $[b,b^{\dag }]=1$. The Landau site $%
|n\rangle $ may be identified with the holomorphic basis $|n\rangle
=(n!)^{-1/2}(b^{\dag })^{n}|0\rangle $. Its wave function is easily
calculable,%
\begin{equation}
\langle \mathbf{x}|n\rangle =\sqrt{\frac{1}{2^{n+1}\pi \ell _{B}^{2}n!}}%
z^{n}e^{-r^{2}/4\ell _{B}^{2}},
\end{equation}%
where $z=\left( x+iy\right) /\ell _{B}$ and $r=|\mathbf{x}|$. We now
construct the Fock space $\mathbb{H}_{\text{LLL}}$ with the use of the
creation operator $c_{\mu }^{\dagger }(n)$ on this Landau site $|n\rangle $,
satisfying $\{c_{\mu }(n),c_{\nu }^{\dagger }(m)\}=\delta _{\mu \nu }\delta
_{nm}$, where $\mu $ denotes the spin index, $\mu =\uparrow $,$\downarrow $.

We expand the two-component electron field $\Psi =(\psi _{\uparrow }$,$\psi
_{\downarrow })$ in terms of wave functions,%
\begin{equation}
\psi _{\mu }(\mathbf{x})=\sum_{n}\langle \mathbf{x}|n\rangle c_{\mu }(n).
\label{ElectInLLL}
\end{equation}%
The physical variables are the number density $\rho (\mathbf{x})=\Psi ^{\dag
}(\mathbf{x})\Psi (\mathbf{x})$ and the spin density $S_{a}(\mathbf{x})=%
\frac{1}{2}\Psi ^{\dag }(\mathbf{x})\tau _{a}\Psi (\mathbf{x})$ with the
Pauli matrix $\tau _{a}$. Substituting (\ref{ElectInLLL}) into these density
operators, we obtain 
\begin{subequations}
\label{PhysiDensi}
\begin{align}
\rho (\mathbf{x})& =\sum \langle m|\mathbf{x}\rangle \langle \mathbf{x}%
|n\rangle \rho (n,m), \\
S_{a}(\mathbf{x})& =\sum \langle m|\mathbf{x}\rangle \langle \mathbf{x}%
|n\rangle S_{a}(n,m)
\end{align}%
with 
\end{subequations}
\begin{subequations}
\label{ProjeDensiMN}
\begin{align}
\rho (n,m)=& \sum_{\mu }c_{\mu }^{\dagger }(m)c_{\mu }(n), \\
S_{a}(n,m)=& \frac{1}{2}\sum_{\mu \nu }c_{\mu }^{\dagger }(m)\left( \tau
_{a}\right) _{\mu \nu }c_{\nu }(n).
\end{align}%
We employ the relation 
\end{subequations}
\begin{equation}
\int \!d^{2}x\,e^{-i\mathbf{qx}}\langle m|\mathbf{x}\rangle \langle \mathbf{x%
}|n\rangle =e^{-\ell _{B}^{2}\mathbf{q}^{2}/4}\langle m|e^{-i\mathbf{qX}%
}|n\rangle ,
\end{equation}%
and transform them in the momentum space, 
\begin{equation}
\rho (\mathbf{q})=e^{-\ell _{B}^{2}\mathbf{q}^{2}/4}\hat{\rho}(\mathbf{q}%
),\quad S_{a}(\mathbf{q})=e^{-\ell _{B}^{2}\mathbf{q}^{2}/4}\hat{S}_{a}(%
\mathbf{q}),  \label{PhysiAndHat}
\end{equation}%
where 
\begin{subequations}
\label{ProjeDensiQ}
\begin{align}
\hat{\rho}(\mathbf{q})=& {\frac{1}{2\pi }}\sum_{mn}\langle m|e^{-i\mathbf{qX}%
}|n\rangle \rho (n,m), \\
\hat{S}_{a}(\mathbf{q})=& {\frac{1}{2\pi }}\sum_{mn}\langle m|e^{-i\mathbf{qX%
}}|n\rangle S_{a}(n,m),
\end{align}%
We call $\hat{\rho}(\mathbf{q})$ and $\hat{S}_{a}(\mathbf{q})$ the bare
densities. In the coordinate space we have the relation 
\end{subequations}
\begin{equation}
\rho (\mathbf{x})=\frac{1}{\pi \ell _{B}^{2}}\int \!d^{2}x^{\prime }\,e^{-|%
\mathbf{x}-\mathbf{x}^{\prime }|^{2}/\ell _{B}^{2}}\hat{\rho}(\mathbf{x}%
^{\prime })  \label{PhysiBareDensi}
\end{equation}%
between the physical density $\rho (\mathbf{x})$ and the bare density $\hat{%
\rho}(\mathbf{x})$.

The density operators generate the SU(2) extension of the W$_{\infty }$
algebra,%
\begin{align}
\lbrack \rho (m,n),\rho (i,j)]& =\delta _{mj}\rho (i,n)-\delta _{in}\rho
(m,j),  \notag \\
\lbrack \rho (m,n),S_{a}(i,j)]& =\delta _{mj}S_{a}(i,n)-\delta
_{in}S_{a}(m,j),  \notag \\
\lbrack S_{a}(m,n),S_{b}(i,j)]& =\frac{i}{2}\varepsilon _{abc}\left[ \delta
_{mj}S_{c}(i,n)+\delta _{in}S_{c}(m,j)\right]  \notag \\
& \hspace{0.7cm}+\frac{1}{4}\delta _{ab}\left[ \delta _{mj}\rho (i,n)-\delta
_{in}\rho (m,j)\right] ,  \label{WalgebP}
\end{align}%
which we have named\cite{EzawaX03B} the W$_{\infty }$(2) algebra. The
electron density and the spin density are intrinsically entangled in the
algebra, as implies that a spin rotation induces a density modulation.

\section{Hamiltonian}

\label{SecHamil}

The Hamiltonian of the QH system consists of a four-fermion interaction term
and the Zeeman term, $H=H_{\text{V}}+H_{\text{Z}}$ with 
\begin{subequations}
\label{HamilQHF}
\begin{align}
H_{\text{V}}& ={\frac{1}{2}}\int \!d^{2}xd^{2}y\,V(\mathbf{x}-\mathbf{y}%
)\delta \rho (\mathbf{x})\delta \rho (\mathbf{y}),  \label{SpinCoulo} \\
H_{\text{Z}}& =-\Delta _{\text{Z}}\int \!d^{2}x\,S_{z}(\mathbf{x}),
\label{SpinZeema}
\end{align}%
where $\delta \rho (\mathbf{x})$ is the density excitation operator, 
\end{subequations}
\begin{equation}
\delta \rho (\mathbf{x})=\rho (\mathbf{x})-\rho _{0},
\end{equation}%
and $\Delta _{Z}=|g|\mu _{B}B_{\perp }$ is the Zeeman gap with $\mu _{B}$
the Bohr magneton. In the actual system $V(\mathbf{x})$ is given by the
Coulomb potential, but here we only assume that it represents a repulsive
interaction.

Substituting the density operators (\ref{PhysiAndHat}) into these we obtain 
\begin{subequations}
\begin{align}
H_{\text{V}}=& \sum_{mnij}{V}_{mnij}\sum_{\mu \nu }c_{\mu }^{\dagger
}(m)c_{\nu }^{\dagger }(i)c_{\nu }(j)c_{\mu }(n)  \label{HamilMonoNO} \notag\\
& +(N_{\Phi }+\delta N)\epsilon _{\text{X}}-\left( N_{\Phi }+2\delta
N\right) \epsilon _{\text{D}}, \\
H_{\text{Z}}=& -\Delta _{\text{Z}}\sum_{n}S_{z}(n,n),
\end{align}%
where 
\end{subequations}
\begin{equation}
{V}_{mnij}={\frac{1}{4\pi }}\int \!d^{2}k\,V(\mathbf{k})e^{-\ell _{B}^{2}%
\mathbf{k}^{2}/2}\langle m|e^{i\mathbf{Xk}}|n\rangle \langle i|e^{-i\mathbf{%
Xk}}|j\rangle ,  \label{Vmnij}
\end{equation}%
with $V(\mathbf{k})$ the Fourier transformation of the potential $V(\mathbf{x%
})$. We have used the notation 
\begin{subequations}
\begin{align}
& \delta N=\int \!d^{2}x\,\delta \rho (\mathbf{x}),  \label{SpinNumbe} \\
& \epsilon _{\text{D}}=\sum_{j}{V}_{nnjj}=\frac{\rho _{0}}{2}\int
\!d^{2}x\,V(\mathbf{x}),  \label{EnergD} \\
& \epsilon _{\text{X}}=\sum_{j}{V}_{njjn}=\frac{\rho _{0}}{2}\ell
_{B}^{2}\int \!d^{2}k\,V(\mathbf{k})e^{-\ell _{B}^{2}\mathbf{k}^{2}/2},
\label{EnergX}
\end{align}%
where $\epsilon _{\text{D}}$ and $\epsilon _{\text{X}}$ are the direct and
exchange energy parameters, respectively.

In the QH state at $\nu =1$, the number of electrons is equal to the number
of Landau sites. One Landau site contains one electron due to the repulsive
interaction. The system without the Zeeman effect ($\Delta _{\text{Z}}=0$)
is most interesting. An intriguing feature is that the SU(2) symmetry is
spontaneously broken by the repulsive interaction and all spins are
spontaneously polarized, as demonstrated in Appendix \ref{AppSpontSymme}.
Thus the QH system is a ferromagnet, where quantum coherence develops
spontaneously.

Without loss of generality we may take

\end{subequations}
\begin{equation}
|\text{g}\rangle =\prod\limits_{n}c_{\uparrow }^{\dag }(n)|0\rangle
\label{GrounState}
\end{equation}%
as the ground state. It satisfies 
\begin{equation}
\rho (m,n)|\text{g}\rangle =\delta _{mn}|\text{g}\rangle ,\quad S_{z}(m,n)|%
\text{g}\rangle =\frac{1}{2}\delta _{mn}|\text{g}\rangle ,
\end{equation}%
and%
\begin{equation}
H_{\text{V}}|\text{g}\rangle =0.
\end{equation}%
A hole-excited and electron-excited states are given by%
\begin{equation}
|\text{h}\rangle =c_{\uparrow }(0)|\text{g}\rangle ,\quad |\text{e}\rangle
=c_{\downarrow }^{\dag }(0)|\text{g}\rangle ,
\end{equation}%
with their energies%
\begin{equation}
\langle \text{h}|H_{\text{V}}|\text{h}\rangle =\langle \text{e}|H_{\text{V}}|%
\text{e}\rangle =\epsilon _{\text{X}}.
\end{equation}%
The main question is whether there are states having the same electron
number but a lower excitation energy compared with the hole-excited or the
electron-excited state.

\section{Microscopic Skyrmion States}

\label{SecMicroSkyrm}

We consider a W$_{\infty }$($2$)-rotated state of the hole-excited state $|$h%
$\rangle $ and the electron-excited state $|$e$\rangle $, 
\begin{subequations}
\label{SkyrmFormuA}
\begin{align}
|\mathfrak{S}_{\text{sky}}^{-}\rangle & =e^{iW^{-}}c_{\uparrow }(0)|\text{g}%
\rangle ,  \label{SkyrmFormuAh} \\
|\mathfrak{S}_{\text{sky}}^{+}\rangle & =e^{iW^{+}}c_{\downarrow }^{\dag
}(0)|\text{g}\rangle ,  \label{SkyrmFormuAe}
\end{align}%
where $W^{\pm }$ are elements of the W$_{\infty }$(2) algebra (\ref{WalgebP}%
). The Coulomb energy of the state depends on $W^{\pm }$, since the W$%
_{\infty }$($2$) rotation modulates not only the spin texture but also the
electron density around the hole or electron excitation. The element $W^{\pm
}$ is to be determined by requiring the excitation energy to be minimized.
As we shall see soon, $|\mathfrak{S}_{\text{sky}}^{-}\rangle $ and $|%
\mathfrak{S}_{\text{sky}}^{+}\rangle $\ describe the skyrmion and
antiskyrmion states. We use the index ($-$) for the skyrmion and ($+$) for
the antiskyrmion.

Similarly, 
\end{subequations}
\begin{align}
|\mathfrak{S}_{\text{sky}}^{-};k\rangle =& e^{iW}c_{\uparrow }(0)\cdots
c_{\uparrow }(k-1)|\text{g}\rangle , \\
|\mathfrak{S}_{\text{sky}}^{+};k\rangle =& e^{i\tilde{W}}c_{\downarrow
}^{\dag }(0)\cdots c_{\downarrow }^{\dag }(k-1)|\text{g}\rangle
\end{align}%
describe a multi-skyrmion and a multi-antiskyrmion as W$_{\infty }$%
(2)-rotated states of a multi-hole state and a multi-electron state.

More generally we may consider a wide class of states presented by%
\begin{equation}
|\mathfrak{S}\rangle =e^{iW}|\mathfrak{S}_{0}\rangle ,  \label{GenerState}
\end{equation}%
where $|\mathfrak{S}_{0}\rangle $ is a state of the form 
\begin{equation}
|\mathfrak{S}_{0}\rangle =\prod_{\mu n}\left[ c_{\mu }^{\dag }\left(
n\right) \right] ^{\nu _{\mu }(n)}|0\rangle ,  \label{PureState}
\end{equation}%
where $\nu _{\mu }(n)$ may take the value either $0$ or $1$ depending
whether the spin state $\mu $ at a site $n$ is occupied or not. The class of
states (\ref{GenerState}) is quite general though it may not embrace all
possible ones. Nevertheless all physically relevant states seem to fall in
this category. Indeed, as far as we know, perturbative excitations are spin
waves and nonperturbative excitations are skyrmions in QH systems. The
corresponding states belong surely to this category.

The electron number of the W$_{\infty }$(2)-rotated state (\ref{GenerState})
is easily calculable,%
\begin{equation}
\langle \mathfrak{S}|N|\mathfrak{S}\rangle =\langle \mathfrak{S}%
_{0}|e^{-iW}Ne^{+iW}|\mathfrak{S}_{0}\rangle =\langle \mathfrak{S}_{0}|N|%
\mathfrak{S}_{0}\rangle ,
\end{equation}%
since $N=\sum_{n}\sum_{\mu }c_{\mu }^{\dagger }(n)c_{\mu }(n)$ is a Casimir
operator. We set%
\begin{equation}
\delta N^{\text{cl}}=\langle \mathfrak{S}|N|\mathfrak{S}\rangle -\langle 
\text{g}|N|\text{g}\rangle .  \label{ElectNumbeA}
\end{equation}%
This is the electron number carried by the excitation described by the state 
$|\mathfrak{S}\rangle $. It is an integer since the electron number of the
state $|\mathfrak{S}\rangle $ is the same as that of the state $|\mathfrak{S}%
_{0}\rangle $.

We examine more in detail the state (\ref{SkyrmFormuA}) with $\delta N^{%
\text{cl}}=\mp 1$. A simplest W$_{\infty }$(2) rotation mixes only
neighboring two sites and is given by the choice of $W^{-}=\sum_{n=0}^{%
\infty }W_{n}^{-}$ with%
\begin{equation}
iW_{n}^{-}=\alpha _{n}\left[ c_{\downarrow }^{\dag }(n)c_{\uparrow
}(n+1)-c_{\uparrow }^{\dag }(n+1)c_{\downarrow }(n)\right] ,
\end{equation}%
where $\alpha _{n}$ is a real parameter. Note that $W_{n}^{-}$ is a
Hermitian operator belonging to the W$_{\infty }$(2) algebra, and $%
[W_{n}^{-},W_{m}^{-}]=0$. We find%
\begin{equation}
e^{+iW^{-}}c_{\uparrow }^{\dag }(n+1)e^{-iW^{-}}=e^{+iW_{n}^{-}}c_{\uparrow
}^{\dag }(n+1)e^{-iW_{n}^{-}}\equiv \xi ^{\dagger }(n),  \label{WtransA}
\end{equation}%
since the spin-up operator $c_{\uparrow }^{\dag }(n+1)$ is contained only in 
$W_{n}^{-}$. We calculate $\xi ^{\dagger }(n)$ using the standard technique
of deriving the differential equation with respect to $\alpha _{n}$. Since
it satisfies%
\begin{equation}
\frac{d^{2}\xi ^{\dagger }(n)}{d\alpha _{n}^{2}}=-\xi ^{\dagger }(n)
\end{equation}%
together with the initial condition $d\xi ^{\dagger }/d\alpha
_{n}=c_{\downarrow }^{\dag }(n)$ at $\alpha _{n}=0$, we integrate it as 
\begin{equation}
\xi ^{\dagger }(n)=u_{-}(n)c_{\downarrow }^{\dag }(n)+v_{-}(n)c_{\uparrow
}^{\dag }(n+1),
\end{equation}%
where we have set $u_{-}(n)=\sin \alpha _{n}$ and $v_{-}(n)=\cos \alpha _{n}$%
. Thus the constraint 
\begin{equation}
u_{-}^{2}(n)+v_{-}^{2}(n)=1  \label{CondiOnU}
\end{equation}%
is imposed on $u_{-}(n)$ and $v_{-}(n)$. We now find%
\begin{equation}
|\mathfrak{S}_{\text{sky}}^{-}\rangle
=e^{iW^{-}}\prod\limits_{n=0}c_{\uparrow }^{\dagger }(n+1)|0\rangle
=\prod\limits_{n=0}\xi ^{\dagger }(n)|0\rangle ,  \label{MicroSkyrmState}
\end{equation}%
where we have used (\ref{WtransA}) and $e^{-iW_{n}^{-}}|0\rangle =0$.

Similarly we choose $W^{+}=\sum_{n}W_{n}^{+}$ in (\ref{SkyrmFormuAe}) with 
\begin{equation}
iW_{n}^{+}=\sum_{n}\beta _{n}\left[ c_{\downarrow }^{\dag }(n+1)c_{\uparrow
}(n)-c_{\uparrow }^{\dag }(n)c_{\downarrow }(n+1)\right] .
\end{equation}%
We find%
\begin{equation}
|\mathfrak{S}_{\text{sky}}^{+}\rangle =c_{\downarrow }^{\dagger
}(0)\prod_{n=0}\zeta ^{\dagger }(n)|0\rangle ,  \label{MicroSkyrmStateAnti}
\end{equation}%
where%
\begin{equation}
\zeta ^{\dagger }(n)=u_{+}(n)c_{\uparrow }^{\dagger
}(n)+v_{+}(n)c_{\downarrow }^{\dagger }(n+1)
\end{equation}%
together with%
\begin{equation}
u_{+}^{2}(n)+v_{+}^{2}(n)=1,  \label{CondiOnV}
\end{equation}%
since $u_{+}(n)=\cos \beta _{n}$ and $v_{+}(n)=\sin \beta _{n}$. The states (%
\ref{MicroSkyrmState}) and (\ref{MicroSkyrmStateAnti}) agree with the
skyrmion state and the antiskyrmion state proposed by Fertig et al.\cite%
{Fertig94B}, respectively.

The operators $\xi (m)$ and $\zeta (n)$ satisfy the standard canonical
anticommutation relations,%
\begin{align}
\{\xi (m),\xi ^{\dagger }(n)\}=& \delta _{mn},\quad \{\xi (m),\xi (n)\}=0, 
\notag \\
\{\zeta (m),\zeta ^{\dagger }(n)\}=& \delta _{mn},\quad \{\zeta (m),\zeta
(n)\}=0,
\end{align}%
with the use of the constrains (\ref{CondiOnU}) and (\ref{CondiOnV}). Since
these states should approach the ground state asymptotically it is necessary
that%
\begin{align}
\lim_{n\rightarrow \infty }u_{-}(n)=& 0,\qquad \lim_{n\rightarrow \infty
}v_{-}(n)=1,  \notag \\
\lim_{n\rightarrow \infty }u_{+}(n)=& 1,\qquad \lim_{n\rightarrow \infty
}v_{+}(n)=0.  \label{CondiUVasymp}
\end{align}%
For later convenience we define%
\begin{align}
u_{-}(-1)=& 1,\qquad v_{-}(-1)=0,  \notag \\
u_{+}(-1)=& 0,\qquad v_{+}(-1)=1  \label{CondiUVatM}
\end{align}%
in accordance with (\ref{CondiOnU}) and (\ref{CondiOnV}).

In what follows we calculate explicitly the classical densities%
\begin{align}
\rho ^{\pm \text{cl}}(m,n)\equiv & \langle \mathfrak{S}_{\text{sky}}^{\pm
}|\rho ^{\pm }(m,n)|\mathfrak{S}_{\text{sky}}^{\pm }\rangle \text{,}
\label{ClassDensi} \\
S_{a}^{\pm \text{cl}}(m,n)\equiv & \langle \mathfrak{S}_{\text{sky}}^{\pm
}|S_{a}^{\pm }(m,n)|\mathfrak{S}_{\text{sky}}^{\pm }\rangle \text{.}
\end{align}%
The basic relations for the skyrmion state are%
\begin{align}
c_{\uparrow }(n)|\mathfrak{S}_{\text{sky}}^{-}\rangle =& v_{-}(n-1)\xi (n-1)|%
\mathfrak{S}_{\text{sky}}^{-}\rangle ,  \notag \\
c_{\downarrow }(n)|\mathfrak{S}_{\text{sky}}^{-}\rangle =& u_{-}(n)\xi (n)|%
\mathfrak{S}_{\text{sky}}^{-}\rangle .  \label{BasicFormuSky}
\end{align}%
They reduce the action of $c$ operators to that of $\xi $, thus allowing to
carry out exact calculus. We employ $\langle \mathfrak{S}_{\text{sky}%
}^{-}|\xi ^{\dag }(m)\xi (n)|\mathfrak{S}_{\text{sky}}^{-}\rangle =\delta
_{mn}$ together with (\ref{BasicFormuSky}) and its conjugate. In this way we
come to%
\begin{align}
\rho ^{\pm }(n,n)& =u_{\pm }^{2}(n)+v_{\pm }^{2}(n-1),  \notag \\
S_{z}^{\pm }(n,n)& =\pm \frac{1}{2}\left[ u_{\pm }^{2}(n)-v_{\pm }^{2}(n-1)%
\right] ,  \notag \\
S_{x}^{\pm }(n+1,n)& =S_{x}^{\pm }(n,n+1)=\frac{1}{2}u_{\pm }(n)v_{\pm }(n),
\notag \\
S_{y}^{\pm }(n+1,n)& =-S_{y}^{\pm }(n,n+1)=\frac{i}{2}u_{\pm }(n)v_{\pm }(n).
\label{MicroDensiA}
\end{align}%
All other components, $\rho ^{\pm }(n,m)$, etc., vanish. Here and hereafter
we omit the superscript "cl". By substituting these into the physical
density (\ref{PhysiDensi}), we obtain 
\begin{align}
\frac{\rho ^{\pm }(\mathbf{x})}{\rho _{0}}=& e^{-|z|^{2}/2}\sum_{n=0}\left[
u_{\pm }^{2}(n)+v_{\pm }^{2}(n-1)\right] \left( \frac{|z|^{2}}{2}\right)
^{n},  \notag \\
\frac{S_{z}^{\pm }(\mathbf{x})}{\rho _{0}}=& \pm \frac{1}{2}%
e^{-|z|^{2}/2}\sum_{n=0}\frac{u_{\pm }^{2}(n)-v_{\pm }^{2}(n-1)}{n!}\left( 
\frac{|z|^{2}}{2}\right) ^{n},  \notag \\
\frac{S_{x}^{\pm }(\mathbf{x})}{\rho _{0}}=& \frac{x}{\sqrt{2}\ell _{B}}%
e^{-|z|^{2}/2}\sum_{n=0}\frac{u_{\pm }(n)v_{\pm }(n)}{n!\sqrt{n+1}}\left( 
\frac{|z|^{2}}{2}\right) ^{n},  \notag \\
\frac{S_{y}^{\pm }(\mathbf{x})}{\rho _{0}}=& \pm \frac{y}{\sqrt{2}\ell _{B}}%
e^{-|z|^{2}/2}\sum_{n=0}\frac{u_{\pm }(n)v_{\pm }(n)}{n!\sqrt{n+1}}\left( 
\frac{|z|^{2}}{2}\right) ^{n}.  \label{MicroDensiC}
\end{align}%
We have calculated the classical densities of the W$_{\infty }$($2$)-rotated
states of a hole-excited and electron-excited states, which contain
infinitely many variables $u_{\pm }(n)$ and $v_{\pm }(n)$. We estimate the
energies of these states and minimize them in later sections.

\section{N-Body Wave Functions}

\label{SecWaveFunct}

We consider the $N$-electron system over $N+1$ sites. The wave function of a
many-body state $|\mathfrak{S}\rangle $ is defined by 
\begin{equation}
\mathfrak{S}_{\mu _{1}\mu _{2}\cdots \mu _{N}}[\mathbf{x}]=\langle 0|\psi
_{\mu _{1}}(\mathbf{x}_{1})\psi _{\mu _{2}}(\mathbf{x}_{2})\cdots \psi _{\mu
_{N}}(\mathbf{x}_{N})|\mathfrak{S}\rangle ,
\end{equation}%
where $\psi _{\mu }(\mathbf{x})$ is given by (\ref{ElectInLLL}), or 
\begin{equation}
\psi _{\mu }(\mathbf{x})=\rho _{0}^{1/2}e^{-|z|^{2}/4}\sum_{n=0}^{N}\alpha
(n)z^{n}c_{\mu }(n)
\end{equation}%
with%
\begin{equation}
\alpha (n)=\sqrt{\frac{1}{2^{n}n!}}.  \label{ParamAlpha}
\end{equation}%
For the hole-excited state $|$h$\rangle =c_{\uparrow }(0)|$g$\rangle $ it is
easy to see 
\begin{equation}
\mathfrak{S}_{\text{h}}[\mathbf{x}]=\prod_{r}^{N}\left( 
\begin{array}{@{\,}c}
z_{r} \\ 
0%
\end{array}%
\right) \mathfrak{S}_{\text{LN}}[\mathbf{x}],
\end{equation}%
where $\mathfrak{S}_{\text{LN}}[\mathbf{x}]$ is the Slater determinant of
the one-body wave functions, 
\begin{equation}
\mathfrak{S}_{\text{LN}}[\mathbf{x}]=\rho _{0}^{N/2}\left\vert 
\begin{array}{cccc}
1 & z_{1} & \cdots & z_{1}^{N-1} \\ 
1 & z_{2} & \cdots & z_{2}^{N-1} \\ 
\vdots & \vdots & \ddots & \vdots \\ 
1 & z_{N} & \cdots & z_{N}^{N-1}%
\end{array}%
\right\vert e^{-\sum_{r=1}^{N}|z_{r}|^{2}/4}.  \label{LaughWave}
\end{equation}%
The wave function of the state $|$h$\rangle $ is factorizable in the
electron coordinates apart from the factor $\mathfrak{S}_{\text{LN}}[\mathbf{%
x}]$.

We proceed to derive the wave function of the skyrmion state (\ref%
{MicroSkyrmState}). First we examine the component with all spins up, to
which only the term $v_{-}(n)c_{\uparrow }^{\dagger }(n+1)$ in $\xi
^{\dagger }(n)$ contributes,%
\begin{equation}
\mathfrak{S}_{\uparrow \uparrow \cdots \uparrow }^{-}[\mathbf{x}%
]=\prod_{n=0}^{N-1}v_{-}(n)\langle 0|\prod_{i=1}^{N}\psi _{\uparrow }(%
\mathbf{x}_{i})\prod_{n=1}^{N}c_{\uparrow }^{\dagger }(n)|0\rangle .
\end{equation}%
Apart from the factor $\prod_{n=0}^{N-1}v_{-}(n)$ this is nothing but the
wave function of a hole. Thus%
\begin{equation}
\mathfrak{S}_{\uparrow \uparrow \cdots \uparrow }^{-}[\mathbf{x}%
]=C_{N}\prod_{n=1}^{N}z_{n}\mathfrak{S}_{\text{LN}}[\mathbf{x}],
\label{SkyrmWaveU}
\end{equation}%
where $C_{N}=\prod_{n=1}^{N}\alpha (n)v_{-}(n-1)$. Next we consider the
component with all spins down, which arises only from the term $%
u_{-}(n)c_{\downarrow }^{\dagger }(n)$ in $\xi ^{\dagger }(n)$,%
\begin{align}
\mathfrak{S}_{\downarrow \downarrow \cdots \downarrow }^{-}[\mathbf{x}]=&
\prod_{n=1}^{N}u_{-}(n)\langle 0|\prod_{i=1}^{N}\psi _{\downarrow }(\mathbf{x%
}_{i})\prod_{n=1}^{N}c_{\downarrow }^{\dagger }(n)|0\rangle  \notag \\
=& C_{N}\prod_{n=0}^{N-1}\beta _{n}\mathfrak{S}_{\text{LN}}[\mathbf{x}].
\label{SkyrmWaveD}
\end{align}%
Here we have set%
\begin{equation}
\beta _{n}=\frac{\alpha (n)u_{-}(n)}{\alpha (n+1)v_{-}(n)}=\frac{u_{-}(n)}{%
v_{-}(n)}\sqrt{2(n+1)}  \label{CondiOnUVa}
\end{equation}%
with (\ref{ParamAlpha}). Comparing (\ref{SkyrmWaveU}) and (\ref{SkyrmWaveD})
we remark that the wave function $\mathfrak{S}_{\downarrow \downarrow \cdots
\downarrow }^{\text{sky}}[\mathbf{x}]$ is obtained by replacing $z_{n}$ with 
$\beta _{n-1}$ within the factor $\prod z_{n}$ of the wave function $%
\mathfrak{S}_{\uparrow \uparrow \cdots \uparrow }[\mathbf{x}]$ for all $n$.
The wave function with mixed spin components is similarly derived, where $%
z_{n}$ is replaced with $\beta _{n-1}$ for certain indices $n$ within the
factor $\prod z_{n}$. In general we derive%
\begin{align}
& \mathfrak{S}_{\mu _{1}\mu _{2}\cdots \mu _{N}}^{-}[\mathbf{x}%
]=C_{N}e^{-\sum_{r=1}^{N}|z_{r}|^{2}/4}  \notag \\
& \times \left\vert 
\begin{array}{cccc}
\left( 
\begin{array}{c}
z_{1} \\ 
\beta _{0}%
\end{array}%
\right) _{\mu _{1}} & z_{1}\left( 
\begin{array}{c}
z_{1} \\ 
\beta _{1}%
\end{array}%
\right) _{\mu _{1}} & \cdots & z_{1}^{N-1}\left( 
\begin{array}{c}
z_{1} \\ 
\beta _{N-1}%
\end{array}%
\right) _{\mu _{1}} \\ 
\left( 
\begin{array}{c}
z_{2} \\ 
\beta _{0}%
\end{array}%
\right) _{\mu _{2}} & z_{2}\left( 
\begin{array}{c}
z_{2} \\ 
\beta _{1}%
\end{array}%
\right) _{\mu _{2}} & \cdots & z_{2}^{N-1}\left( 
\begin{array}{c}
z_{2} \\ 
\beta _{N-1}%
\end{array}%
\right) _{\mu _{2}} \\ 
\vdots & \vdots & \ddots & \vdots \\ 
\left( 
\begin{array}{c}
z_{N} \\ 
\beta _{0}%
\end{array}%
\right) _{\mu _{N}} & z_{N}\left( 
\begin{array}{c}
z_{N} \\ 
\beta _{1}%
\end{array}%
\right) _{\mu _{N}} & \cdots & z_{N}^{N-1}\left( 
\begin{array}{c}
z_{N} \\ 
\beta _{N-1}%
\end{array}%
\right) _{\mu _{N}}%
\end{array}%
\right\vert .  \label{SkyrmWaveFunct}
\end{align}%
This is the wave function of the skyrmion.

It is notable that, when all $\beta _{n}$ are equal ($\beta _{n}=\sqrt{2}%
\omega $), or%
\begin{equation}
\frac{u_{-}(n)}{v_{-}(n)}\sqrt{2(n+1)}=\beta _{n}=\sqrt{2}\omega ,
\label{FactoCondi}
\end{equation}%
it is reduced to 
\begin{equation}
\mathfrak{S}_{\text{sky}}^{-}[\mathbf{x}]=\prod_{r}^{N}\left( 
\begin{array}{@{\,}c}
z_{r} \\ 
\sqrt{2}\omega%
\end{array}%
\right) \mathfrak{S}_{\text{LN}}[\mathbf{x}].  \label{FactoSkyrmWave}
\end{equation}%
The wave function is factorizable as in the case of a hole. We call such a
skyrmion the factorizable skyrmion. As we shall see in the following
section, this is the wave function of the skyrmion in the hard-core model.
It is quite difficult to write down the wave function of an antiskyrmion in
terms of the analytic variable $z_{r}$. We discuss it in Appendix \ref%
{SecAntiSkyrm}.

\section{Hard-Core Interaction}

\label{SecHardCore}

We first investigate a detailed structure of skyrmions in the system
governed by the hard-core interaction,\cite{LeeX01D}%
\begin{equation}
V(\mathbf{x}-\mathbf{y})=\delta ^{2}(\mathbf{x}-\mathbf{y}),
\label{ContaPoten}
\end{equation}%
instead of the Coulomb interaction. The Hamiltonian (\ref{SpinCoulo}) reads%
\begin{equation}
H_{\text{hc}}={\frac{1}{2}}\int \!d^{2}x\;\delta \rho (\mathbf{x})\delta
\rho (\mathbf{x}).  \label{ContaHamil}
\end{equation}%
All previous formulas hold with%
\begin{align}
{V}_{mnij}=& {\frac{1}{8\pi ^{2}}}\int \!d^{2}k\,e^{-\ell _{B}^{2}\mathbf{k}%
^{2}/2}\langle m|e^{i\mathbf{Xk}}|n\rangle \langle i|e^{-i\mathbf{Xk}%
}|j\rangle  \notag \\
=& \frac{1}{8\pi \ell _{B}^{2}}\frac{\sqrt{(m+i)!(n+j)!}}{\sqrt{m!i!n!j!}}%
\frac{\delta _{m+i,n+j}}{\sqrt{2^{m+i+n+j}}},  \label{ContaStepA}
\end{align}%
and%
\begin{equation}
\epsilon _{\text{D}}=\epsilon _{\text{X}}=\frac{1}{4\pi \ell _{B}^{2}}.
\end{equation}%
We rewrite (\ref{ContaHamil}) into the normal ordered form,%
\begin{equation}
H_{\text{hc}}=\int \!d^{2}x\;\psi _{\uparrow }^{\dagger }(\mathbf{x})\psi
_{\downarrow }^{\dagger }(\mathbf{x})\psi _{\downarrow }(\mathbf{x})\psi
_{\uparrow }(\mathbf{x})-\frac{1}{4\pi \ell _{B}^{2}}\int \!d^{2}x\,\delta
\rho (\mathbf{x}).
\end{equation}%
The ground state $|$g$\rangle $ is given by (\ref{GrounState}) as an
eigenstate of the Hamiltonian, $H_{\text{hc}}|$g$\rangle =0$. A hole-excited
and electron-excited states are given by $|$h$\rangle =c_{\uparrow }(0)|$g$%
\rangle $ and $|$e$\rangle =c_{\downarrow }^{\dag }(0)|$g$\rangle $.

We determine the skyrmion and antiskyrmion states by minimizing the energy
of the W$_{\infty }$($2$)-rotated state (\ref{SkyrmFormuA}). We deal with
the skyrmion state (\ref{MicroSkyrmState}) explicitly, but similar formulas
follow also for the antiskyrmion state (\ref{MicroSkyrmStateAnti}).

We define the state%
\begin{align}
|\mathfrak{S}^{\prime }\rangle =& \psi _{\downarrow }(\mathbf{x})\psi
_{\uparrow }(\mathbf{x})|\mathfrak{S}_{\text{sky}}^{-}\rangle  \notag \\
=& \sum_{mn}\varphi _{m}(\mathbf{x})\varphi _{n}(\mathbf{x})c_{\downarrow
}(m)c_{\uparrow }(n)|\mathfrak{S}_{\text{sky}}^{-}\rangle .
\end{align}%
Using (\ref{BasicFormuSky}) we come to 
\begin{equation}
|\mathfrak{S}^{\prime }\rangle =\sum_{m=0}\sum_{n=0}\varphi _{m}(\mathbf{x}%
)\varphi _{n+1}(\mathbf{x})u_{-}(m)v_{-}(n)\xi (m)\xi (n)|\mathfrak{S}_{%
\text{sky}}^{-}\rangle ,
\end{equation}%
where we have used $v(-1)=0$. Hence 
\begin{equation}
\int \!d^{2}x\;\langle \mathfrak{S}^{\prime }|\mathfrak{S}^{\prime }\rangle
=2\sum_{k,l=0}v_{-}(k)u_{-}(l)v_{-}(l)u_{-}(k)V_{k+1,k,l,l+1}.
\label{StepHCa}
\end{equation}%
Using (\ref{ContaStepA}) we obtain%
\begin{align}
& \int \!d^{2}x\;\langle \mathfrak{S}^{\prime }|\mathfrak{S}^{\prime }\rangle
\\
=& \frac{1}{8\pi \ell _{B}^{2}}\sum_{k,l=0}\frac{(k+l+1)!}{k!l!2^{k+l+1}}%
\left[ \frac{v_{-}(k)u_{-}(l)}{\sqrt{k+1}}-\frac{v_{-}(l)u_{-}(k)}{\sqrt{l+1}%
}\right] ^{2}\geqslant 0.
\end{align}%
The skyrmion energy $\langle \mathfrak{S}_{\text{sky}}^{-}|H_{\text{hc}}|%
\mathfrak{S}_{\text{sky}}^{-}\rangle $ is minimized when the equality is
achieved, as occurs when the factorizable-skyrmion condition (\ref%
{FactoCondi}) is satisfied, or%
\begin{equation}
u_{-}^{2}(n)=\frac{\omega ^{2}}{n+1+\omega ^{2}},\quad v_{-}^{2}(n)=\frac{n+1%
}{n+1+\omega ^{2}}  \label{SkyrmAnzatL}
\end{equation}%
due to the normalization (\ref{CondiOnU}).

It follows $|\mathfrak{S}^{\prime }\rangle =0$ when the equality holds in (%
\ref{StepHCa}). Hence, $|\mathfrak{S}_{\text{sky}}^{-}\rangle $ is an
eigenstate of the Hamiltonian with $\delta N^{\text{cl}}=-1$,%
\begin{equation}
H_{\text{hc}}|\mathfrak{S}_{\text{sky}}^{-}\rangle =\frac{1}{4\pi }\int
\!d^{2}x\,\delta \rho (\mathbf{x})|\mathfrak{S}_{\text{sky}}^{-}\rangle =-%
\frac{1}{4\pi }\delta N^{\text{cl}}|\mathfrak{S}_{\text{sky}}^{-}\rangle .
\label{ContaSkyrmEnerg}
\end{equation}%
Similarly, when%
\begin{equation}
u_{+}^{2}(n)=\frac{n+1}{n+1+\omega ^{2}},\quad v_{+}^{2}(n)=\frac{\omega ^{2}%
}{n+1+\omega ^{2}}
\end{equation}%
the antiskyrmion state is an eigenstate of the Hamiltonian with $\delta N^{%
\text{cl}}=1$,%
\begin{equation}
H_{\text{hc}}|\mathfrak{S}_{\text{sky}}^{+}\rangle =\frac{1}{4\pi }\int
\!d^{2}x\,\delta \rho (\mathbf{x})|\mathfrak{S}_{\text{sky}}^{+}\rangle =%
\frac{1}{4\pi }\delta N^{\text{cl}}|\mathfrak{S}_{\text{sky}}^{+}\rangle .
\end{equation}%
We may summarize them as $H_{\text{hc}}|\mathfrak{S}_{\text{sky}}^{\pm
}\rangle =E_{\text{hc}}|\mathfrak{S}_{\text{sky}}^{\pm }\rangle $ with $E_{%
\text{hc}}=|\delta N^{\text{cl}}|/4\pi $. The skyrmion state and the
antiskyrmion state are the lowest energy states possessing the electron
numbers $\delta N^{\text{cl}}=-1$ and $\delta N^{\text{cl}}=1$,
respectively. In conclusion, we have determined the W$_{\infty }$(2)-rotated
states (\ref{SkyrmFormuA}) by minimizing the energy, though their energy is
independent of the scale $\omega $ and degenerates with the hole-excited and
electron-excited state.

\begin{figure}[h]
\includegraphics[width=0.5\textwidth]{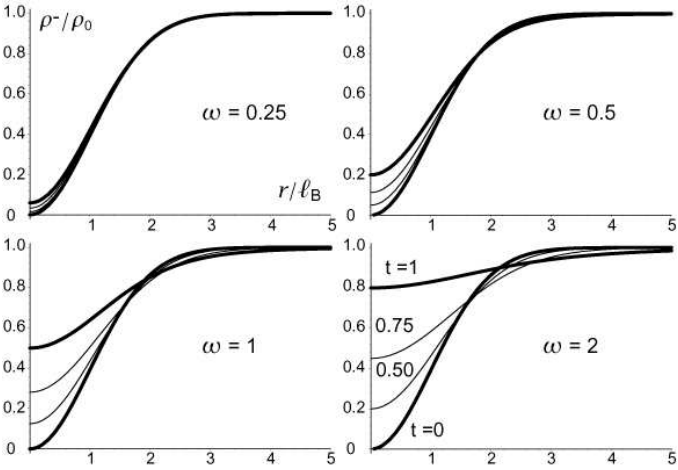}
\caption{The density modulation ($\protect\rho ^{-}/\protect\rho _{0}$)
accompanied with the skyrmion excitation is plotted as a function of the
radius ($r/\ell _{B}$). The heavy curve ($t=1$) is for the hard-core model
based on (\protect\ref{MicroDensiD}) for the choice of $\protect\omega =0.25$%
, $0.5$, $1$, $2$, while the heavy curve ($t=0$) represents the density of a
hole. The thin curves represent the interpolating formula (\protect\ref%
{MicroDensiE}) for $t=0.50$, $0.75$, where the parameter $t$ is defined by (%
\protect\ref{CondiOnUVb}).}
\label{FigSkyDensi}
\end{figure}

The physical densities can be expressed in terms of the Kummer function $%
M(a;b;x)$, 
\begin{equation}
M(a;a+1;x)=a\sum_{n=0}^{\infty }\frac{x^{n}}{(n+a)n!},
\end{equation}%
as 
\begin{align}
\frac{\delta \rho ^{\pm }(\mathbf{x})}{\rho _{0}}=& \pm e^{-\frac{1}{2}%
|z|^{2}}M(\omega ^{2};\omega ^{2}+1;|z|^{2}/2)  \notag \\
& \mp \frac{\omega ^{2}}{\omega ^{2}+1}e^{-\frac{1}{2}z^{2}}M(\omega
^{2}+1;\omega ^{2}+2;|z|^{2}/2),  \notag \\
\frac{S_{z}^{\pm }(\mathbf{x})}{\rho _{0}}=& \frac{1}{2}-\frac{1}{2}e^{-%
\frac{1}{2}|z|^{2}}M(\omega ^{2};\omega ^{2}+1;|z|^{2}/2)  \notag \\
& -\frac{1}{2}\frac{\omega ^{2}}{\omega ^{2}+1}e^{-\frac{1}{2}%
|z|^{2}}M(\omega ^{2}+1;\omega ^{2}+2;|z|^{2}/2),  \notag \\
\frac{S_{x}^{\pm }(\mathbf{x})}{\rho _{0}}=& \frac{1}{\sqrt{2}}\frac{\omega
x/\ell _{B}}{\omega ^{2}+1}e^{-\frac{1}{2}|z|^{2}}M(\omega ^{2}+1;\omega
^{2}+2;|z|^{2}/2),  \notag \\
\frac{S_{y}^{\pm }(\mathbf{x})}{\rho _{0}}=& \frac{\pm 1}{\sqrt{2}}\frac{%
\omega y/\ell _{B}}{\omega ^{2}+1}e^{-\frac{1}{2}|z|^{2}}M(\omega
^{2}+1;\omega ^{2}+2;|z|^{2}/2),  \label{MicroDensiD}
\end{align}%
with $\left\vert z\right\vert ^{2}=r^{2}/\ell _{B}^{2}$. We have illustrated
the density $\rho ^{-}(r)$ for typical values of the parameters $\omega $ in
Fig.\ref{FigSkyDensi}.

Using the relations%
\begin{align}
M(1;b;x)=& 1+\frac{x}{b}\hspace{0.5mm}M(1;b+1;x),  \notag \\
M(a;b;z)=& e^{z}M(b-a;b;-z),  \label{SkyrmStepB}
\end{align}%
we are able to summarize the densities as%
\begin{equation}
\delta \rho ^{+}(\mathbf{x})=-\delta \rho ^{-}(\mathbf{x}),\quad S_{a}^{\pm
}(\mathbf{x})=\rho ^{-}(\mathbf{x})\mathit{S}_{a}^{\pm }(\mathbf{x}),
\label{SkyrmConfiConta}
\end{equation}%
with%
\begin{align}
\mathit{S}_{x}(\mathbf{x})=& \frac{2\lambda x}{r^{2}+\lambda ^{2}},\quad 
\mathit{S}_{y}(\mathbf{x})=\frac{\mp 2\lambda y}{r^{2}+\lambda ^{2}},  \notag
\\
\mathit{S}_{z}(\mathbf{x})=& \frac{r^{2}-\lambda ^{2}}{r^{2}+\lambda ^{2}},
\label{StandSkyrm}
\end{align}%
and%
\begin{equation}
\rho ^{-}(\mathbf{x})=\frac{r^{2}+\lambda ^{2}}{2\ell _{B}^{2}+\lambda ^{2}}%
M(1;\omega ^{2}+2;-|z|^{2}/2)\rho _{0},  \label{KummeEqB}
\end{equation}%
where we have set $\lambda =\sqrt{2}\omega \ell $. It is interesting that
the physical spin densities $S_{a}^{\pm }(\mathbf{x})$ are factorizable into
the number density $\rho ^{-}(\mathbf{x})$ and the normalized spin field $%
\mathit{S}_{a}^{\pm }(\mathbf{x})$ as in (\ref{SkyrmConfiConta}). The
normalized spin field (\ref{StandSkyrm}) agrees with the
nonlinear-sigma-model skyrmion configuration\cite{BookRajaraman}.

We explore the properties of the density modulation. The skyrmion is reduced
to a hole for $\lambda =0$, where $\rho ^{-}(\mathbf{x})$ is given by%
\begin{equation}
\rho ^{-}(\mathbf{x})=\left( 1-e^{-r^{2}/2\ell _{B}^{2}}\right) \rho _{0}
\end{equation}%
and vanishes at $r=0$. On the other hand, $\rho ^{-}(\mathbf{x})>0$ for all $%
\lambda \neq 0$ with 
\begin{equation}
\rho ^{-}(0)=\frac{\lambda ^{2}}{\lambda ^{2}+2\ell _{B}^{2}}\rho _{0}.
\end{equation}%
We can expand $\rho ^{-}(\mathbf{x})$ in a power series of $\ell _{B}^{2}$
as follows. Let us set%
\begin{equation}
\rho ^{-}(\mathbf{x})=\left( 1-\frac{2\ell _{B}^{2}}{\lambda ^{2}+2\ell
_{B}^{2}}f(r^{2})\right) \rho _{0}.  \label{KummeEqA}
\end{equation}%
By comparing (\ref{KummeEqA}) with (\ref{KummeEqB}), $f(x)$ is found to be
the Kummer function of the form%
\begin{equation}
f(x)=M(2;\frac{\lambda ^{2}}{2\ell _{B}^{2}}+2;-\frac{x}{\ell _{B}^{2}})
\end{equation}%
with $x=r^{2}$, and satisfies the Kummer equation,%
\begin{equation}
2\ell _{B}^{2}x\frac{d^{2}f}{dx^{2}}+(\lambda ^{2}+x+4\ell _{B}^{2})\frac{df%
}{dx}+2f=0.  \label{KummeEq}
\end{equation}%
This can be solved by expanding $f$ in a power series of $\ell _{B}^{2}$, $%
f=\sum_{n=0}f_{n}\ell _{B}^{2n}$. In particular, the lowest order term is
given by setting $\ell _{B}^{2}=0$, whose result is%
\begin{equation}
f_{0}(x)=\frac{\lambda ^{4}}{\left( x+\lambda ^{2}\right) ^{2}}.
\end{equation}%
Hence we have 
\begin{equation}
\rho ^{-}(\mathbf{x})=\left( 1-\frac{2\lambda ^{2}\ell _{B}^{2}}{\left(
r^{2}+\lambda ^{2}\right) ^{2}}+O(\ell _{B}^{4})\right) \rho _{0},
\label{SkyrmStepD}
\end{equation}%
or 
\begin{subequations}
\label{SkyrmStepE}
\begin{align}
\frac{\delta \rho ^{\pm }(\mathbf{x})}{\rho _{0}}=& \pm \frac{2\lambda
^{2}\ell _{B}^{2}}{\left( r^{2}+\lambda ^{2}\right) ^{2}}+O(\ell _{B}^{4}),
\label{SkyrmStepEa} \\
\frac{S_{z}^{\pm }(\mathbf{x})}{\rho _{0}}=& \frac{1}{2}-\frac{\lambda ^{2}}{%
r^{2}+\lambda ^{2}}+O(\ell _{B}^{2}).  \label{SkyrmStepEb}
\end{align}%
The densities $\rho ^{\pm }(\mathbf{x})$ as well as $S_{z}^{\pm }(\mathbf{x})
$ approach the ground-state values only polynomially.

Finally we examine what happens when the Zeeman interaction is taken into
account. The number of spins flipped around a skyrmion is given by 
\end{subequations}
\begin{equation}
N_{\text{spin}}=\int \!d^{2}x\;\left\{ S_{z}^{\text{cl}}(\mathbf{x})-\frac{1%
}{2}\rho _{0}\right\} ,
\end{equation}%
which we call the skyrmion spin. Substituting (\ref{SkyrmConfiConta}) into
this, unless $\lambda =0$, we find $N_{\text{spin}}^{\text{sky}}$ to diverge
logarithmically due to the asymptotic behavior,%
\begin{equation}
\lim_{r\rightarrow \infty }S_{z}^{\text{cl}}(\mathbf{x})=\frac{\rho _{0}}{2}%
\left( 1-2\frac{\lambda ^{2}}{r^{2}}\right) .  \label{AsympSz}
\end{equation}%
The Zeeman energy $H_{\text{Z}}^{\text{sky}}=-\Delta _{\text{Z}}N_{\text{spin%
}}$ is divergent, except for the hole, from the infrared contribution
however small the Zeeman effect is. The factorizable skyrmion (\ref%
{SkyrmConfiConta}) is no longer valid. There exists a skyrmion state which
has a finite Zeeman energy: See an example of (\ref{CondiOnUVb}) we use for
the Coulomb interaction. Nevertheless, we can show that the hole state has
the lowest energy. The reason reads as follows. The factorizable skyrmion is
an eigenstate of the hard-core Hamiltonian, $H_{\text{hc}}|\mathfrak{S}_{%
\text{sky}}^{\pm }\rangle =E_{\text{hc}}|\mathfrak{S}_{\text{sky}}^{\pm
}\rangle $ with $E_{\text{hc}}=|\delta N^{\text{cl}}|/4\pi $, as in (\ref%
{ContaSkyrmEnerg}). Accordingly any spin texture $|\mathfrak{S}\rangle $
possessing the same electron number $\delta N^{\text{cl}}$ has a higher
energy, $\langle \mathfrak{S}|H_{\text{hc}}|\mathfrak{S}\rangle \geq E_{%
\text{hc}}$. Furthermore its Zeeman energy is larger than that of the hole, $%
\langle \mathfrak{S}|H_{\text{Z}}|\mathfrak{S}\rangle \geq \frac{1}{2}\Delta
_{\text{Z}}$. Hence, 
\begin{equation}
\langle \mathfrak{S}|\left( H_{\text{hc}}+H_{\text{Z}}\right) |\mathfrak{S}%
\rangle \geq E_{\text{hc}}+\frac{1}{2}\Delta _{\text{Z}},
\end{equation}%
where the equality holds for the hole state. Consequently there are no
skyrmions in the presence of the Zeeman interaction in the system with the
hard-core interaction.

\section{Coulomb Interaction}

\label{SecInterCoulo}

We next investigate the realistic system (\ref{SpinCoulo}) governed by the
Coulomb Hamiltonian $H_{\text{C}}$ with the potential%
\begin{equation}
V(\mathbf{x}-\mathbf{y})=\frac{e^{2}}{4\pi \varepsilon |\mathbf{x}-\mathbf{y}%
|},  \label{CouloHamil}
\end{equation}%
for which the exchange energy parameter reads%
\begin{equation}
\epsilon _{\text{X}}=\frac{\sqrt{2\pi }}{4}\epsilon _{\text{C}}
\end{equation}%
with $\epsilon _{\text{C}}=e^{2}/4\pi \varepsilon \ell _{B}$ the Coulomb
energy unit. It is hard to construct the skyrmion state explicitly as an
eigenstate. We are satisfied to estimate the excitation energy of a skyrmion
by minimizing the expectation value of the Hamiltonian.

Before so doing, it is instructive to estimate the Coulomb energy by using
the factorizable skyrmion (\ref{SkyrmConfiConta}) obtained in the hard-core
model. The result is very different from the one in the hard-core model. The
Coulomb energy is a monotonously decreasing function of the scale parameter $%
\lambda $, and%
\begin{equation}
\lim_{\lambda \rightarrow \infty }\langle H_{\text{C}}\rangle =\pm 8\pi J_{s}
\end{equation}%
with%
\begin{equation}
J_{s}={\frac{1}{16\sqrt{2\pi }}}\epsilon _{\text{C}}.  \label{SpinStiff}
\end{equation}%
Consequently, an infinitely large skyrmion is necessarily excited. However,
when the Zeeman interaction is introduced, the Zeeman energy of the
factorizable skyrmion diverges due to the asymptotic behavior (\ref{AsympSz}%
). Namely the factorizable skyrmion cannot be a quasiparticle in the
realistic Coulomb system.

It is necessary to consider a skyrmion not factorizable as in (\ref%
{SkyrmConfiConta}). Making a slight generalization of the parametrization (%
\ref{SkyrmAnzatL}) we search for a skyrmion possessing a finite energy even
in the presence of the Zeeman effect. We propose an anzats, 
\begin{equation}
u_{-}^{2}(n)=v_{+}^{2}(n)=\frac{\omega ^{2}t^{2n+2}}{n+1+\omega ^{2}}.
\label{CondiOnUVb}
\end{equation}%
The parameter $t$ presents a smooth interpolation between the hole ($t=0$)
and the factorizable skyrmion ($t=1$). This anzats automatically satisfies
the condition (\ref{CondiUVatM}) for $n=-1$.

Substituting (\ref{CondiOnUVb}) into (\ref{MicroDensiC}) we obtain%
\begin{align}
\frac{\delta \rho ^{\pm }(\mathbf{x})}{\rho _{0}}=& \pm e^{-\frac{1}{2}%
|z|^{2}}M(\omega ^{2};\omega ^{2}+1;\frac{t|z|^{2}}{2})  \notag \\
& \mp \frac{t^{2}\omega ^{2}}{\omega ^{2}+1}e^{-\frac{1}{2}|z|^{2}}M(\omega
^{2}+1;\omega ^{2}+2;\frac{t|z|^{2}}{2}),  \notag \\
\frac{S_{z}^{\pm }(\mathbf{x})}{\rho _{0}}=& \frac{1}{2}-\frac{1}{2}e^{-%
\frac{1}{2}|z|^{2}}M(\omega ^{2};\omega ^{2}+1;\frac{t|z|^{2}}{2})  \notag \\
& -\frac{1}{2}\frac{t^{2}\omega ^{2}}{\omega ^{2}+1}e^{-\frac{1}{2}%
|z|^{2}}M(\omega ^{2}+1;\omega ^{2}+2;\frac{t|z|^{2}}{2}),  \notag \\
\frac{S_{x}^{\pm }(\mathbf{x})}{\rho _{0}}=& \frac{t\omega x/\ell _{B}}{%
\sqrt{2}}e^{-\frac{1}{2}|z|^{2}}\sum_{n}\frac{1}{n!}\frac{\vartheta _{n}(w,t)%
}{n+1+\omega ^{2}}\left( \frac{t|z|^{2}}{2}\right) ^{n},  \notag \\
\frac{S_{y}^{\pm }(\mathbf{x})}{\rho _{0}}=& \frac{\pm t\omega y/\ell _{B}}{%
\sqrt{2}}e^{-\frac{1}{2}|z|^{2}}\sum_{n}\frac{1}{n!}\frac{\vartheta _{n}(w,t)%
}{n+1+\omega ^{2}}\left( \frac{t|z|^{2}}{2}\right) ^{n},  \label{DensiCoulo}
\end{align}%
where%
\begin{equation}
\vartheta _{n}(w,t)=\sqrt{1+\frac{1-t^{2n+2}}{n+1}\omega ^{2}}.
\label{FunctStepA}
\end{equation}%
It is easy to see that various densities approach the ground-state values
exponentially fast for all $t\neq 1$. Using (\ref{SkyrmStepB}), we can
rewrite these as%
\begin{align}
\frac{\delta \rho ^{\pm }(\mathbf{x})}{\rho _{0}}=& \pm e^{-\frac{1}{2}%
(1-t^{2})|z|^{2}}M(1;\omega ^{2}+1;-t^{2}|z|^{2}/2)  \notag \\
& \mp \frac{t^{2}\omega ^{2}}{\omega ^{2}+1}e^{-\frac{1}{2}%
(1-t^{2})|z|^{2}}M(1;\omega ^{2}+2;-t^{2}|z|^{2}/2),  \notag \\
\frac{S_{z}^{\pm }(\mathbf{x})}{\rho _{0}}=& \frac{1}{2}-\frac{1}{2}e^{-%
\frac{1}{2}(1-t^{2})|z|^{2}}M(1;\omega ^{2}+1;-t^{2}|z|^{2}/2)  \notag \\
& -\frac{1}{2}\frac{t^{2}\omega ^{2}}{\omega ^{2}+1}e^{-\frac{1}{2}%
(1-t^{2})|z|^{2}}M(1;\omega ^{2}+2;-t^{2}|z|^{2}/2).  \label{MicroDensiE}
\end{align}
We have illustrated the density $\rho ^{-}(r)$ for typical values of the
parameters $\omega $ and $t$ in Fig.\ref{FigSkyDensi}.

It is instructive to expand them in power series of $\ell _{B}^{2}$. Setting%
\begin{equation}
(1-t^{2})|z|^{2}=(1-t^{2})r^{2}/\ell _{B}^{2}=r^{2}/\left( \beta ^{2}+\ell
_{B}^{2}\right) ,
\end{equation}%
and writing down similar equations to (\ref{KummeEq}), we obtain the lowest
order term as 
\begin{subequations}
\begin{align}
\frac{\delta \rho ^{\pm }(\mathbf{x})}{\rho _{0}}=& \pm \left[ \frac{%
2\lambda ^{2}\ell _{B}^{2}}{\left( r^{2}+\lambda ^{2}\right) ^{2}}+\frac{%
2\lambda ^{2}\ell _{B}^{2}}{\left( r^{2}+\lambda ^{2}\right) \beta ^{2}}%
+O(\ell _{B}^{4})\right] e^{-r^{2}/\beta ^{2}},  \label{SkyrmStepFa} \\
\frac{S_{z}^{\pm }(\mathbf{x})}{\rho _{0}}=& \frac{1}{2}-\left[ \frac{%
\lambda ^{2}}{r^{2}+\lambda ^{2}}+O(\ell _{B}^{2})\right] e^{-r^{2}/\beta
^{2}}.
\end{align}%
The Zeeman energy remains finite due to a rapid decrease to the ground-state
value.

In calculating the expectation value of the Coulomb Hamiltonian we appeal to
the decomposition formula, 
\end{subequations}
\begin{equation}
\langle H_{\text{C}}\rangle =H_{\text{D}}^{\text{cl}}+H_{\text{X}}^{\text{cl}%
},  \label{DecomFormuA}
\end{equation}%
dictating that the Coulomb energy consists of the direct energy $H_{\text{D}%
}^{\text{cl}}$ and the exchange energy $H_{\text{X}}^{\text{cl}}$. We have
already used it on a case-by-case basis in our previous papers.\cite%
{EzawaX03B,EzawaX04B,EzawaX05B} In Appendix \ref{SecDecomFormu} we derive
the formula for the class of states (\ref{GenerState}). Here the direct and
exchange energies read 
\begin{subequations}
\label{DecomDX}
\begin{align}
H_{\text{D}}^{\text{cl}}=& \pi \int \!d^{2}k\,V(\mathbf{k})e^{-\frac{1}{2}%
\ell _{B}^{2}k^{2}}\delta \hat{\rho}^{\text{cl}}(-\mathbf{k})\delta \hat{\rho%
}^{\text{cl}}(\mathbf{k}),  \label{DecomD} \\
H_{\text{X}}^{\text{cl}}=& \pi \int \!d^{2}k\,\delta V_{\text{X}}(\mathbf{k})%
\left[ \hat{S}_{a}^{^{\text{cl}}}(-\mathbf{k})\hat{S}_{a}^{^{\text{cl}}}(%
\mathbf{k})+\frac{1}{4}\hat{\rho}^{\text{cl}}(-\mathbf{k})\hat{\rho}^{\text{%
cl}}(\mathbf{k})\right] ,  \label{DecomX}
\end{align}%
in terms of the bare densities, where $\delta V_{\text{X}}(\mathbf{k})=V_{%
\text{X}}(0)-V_{\text{X}}(\mathbf{k})$ with 
\end{subequations}
\begin{align}
V_{\text{X}}(\mathbf{k})\equiv & \frac{\ell _{B}^{2}}{\pi }\int
\!d^{2}k\,e^{-i\ell _{B}^{2}\mathbf{k}\wedge \mathbf{k}^{\prime }}e^{-\frac{1%
}{2}\ell _{B}^{2}k^{\prime }{}^{2}}V(\mathbf{k}^{\prime })  \notag \\
=& 4\varepsilon _{\text{X}}\ell _{B}^{2}e^{-\frac{1}{4}\ell
_{B}^{2}k^{2}}I_{0}\left( \frac{k^{2}}{4}\right) ,
\end{align}%
and $I_{0}(z)$ is the modified Bessel function.

\begin{figure}[h]
\includegraphics[width=0.4\textwidth]{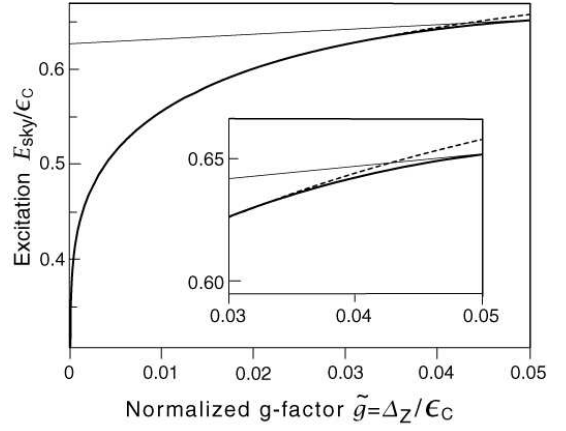}
\caption{The skyrmion excitation energy is plotted as a function of the
normalized Zeeman gap $\widetilde{g}=\Delta _{\text{Z}}/\protect\epsilon _{%
\text{C}}$. The heavy solid curve is obtained by the numerical analysis
based on the density formulas (\protect\ref{DensiCoulo}). The thin line is
the hole excitation energy. It is seen that a skyrmion is excited for $%
\widetilde{g}<0.045$ but a hole is practically excited for $\widetilde{g}%
>0.045$. We have also given the excitation energy calculated analytically
based on the simplified formula (\protect\ref{SimplResul}), which is
represented by the dotted curve. The two curves are practically identical
for $\widetilde{g}<0.038.$ The inset is an enlargement of the figure near $%
\widetilde{g}=0.04$, where three curves meet.}
\label{FigSkyEneSpinTh}
\end{figure}

We are able to determine the parameters $\omega $ and $t$ so as to minimize
the sum of the Coulomb and Zeeman energies, $\langle H\rangle _{\text{sky}%
}=\langle H_{\text{C}}\rangle _{\text{sky}}+\langle H_{\text{Z}}\rangle _{%
\text{sky}}$, as a function of the Zeeman gap $\Delta _{\text{Z}}$.
Calculating numerically $\langle H\rangle _{\text{sky}}$ as a function of $%
\omega $ and $t$ for a given value of $\Delta _{\text{Z}}$, we determine the
values of $\omega $ and $t$ which minimizes $\langle H\rangle _{\text{sky}}$%
. In this way we obtain the skyrmion excitation energy $\langle H\rangle _{%
\text{sky}}$ as a function of $\Delta _{\text{Z}}$. We have plotted the
excitation energy $\langle H\rangle _{\text{sky}}$ as a function of the
normalized Zeeman gap $\widetilde{g}=\Delta _{\text{Z}}/\epsilon _{\text{C}}$
in Fig.\ref{FigSkyEneSpinTh}, where $\epsilon _{\text{C}}$ is the Coulomb
unit.

\section{Experimental Data}

\label{SecExperEvide}

A skyrmion excitation is characterized by a coherent excitation of spins.
Hence an evidence of the skyrmion excitation is given if $N_{\text{spin}%
}>1/2 $, where $N_{\text{spin}}=1/2$ for a hole or an electron excitation.
It is a remarkable fact that skyrmions have already been observed
experimentally in QH systems\cite{Barrett95L,Schmeller95L,Aifer96L}.

Let us show the number of flipped spins is given by%
\begin{equation}
N_{\text{spin}}=\frac{d\langle H\rangle _{\text{sky}}}{d\Delta _{\text{Z}}}.
\label{SpinNumbeEnerg}
\end{equation}%
The excitation energy $E_{\text{sky}}=\langle H\rangle _{\text{sky}}$ is the
sum of the Coulomb energy $E_{\text{C}}=\langle H_{\text{C}}\rangle _{\text{%
sky}}$ and Zeeman energy $\Delta _{\text{Z}}N_{\text{spin}}$, which depend
on a set of parameters ($t$,$\omega $) denoted collectively by $t$ for
simplicity. The quantity to minimize with respect to $t$ is 
\begin{equation}
E_{\text{sky}}(t)=E_{\text{C}}(t)+\Delta _{\text{Z}}N_{\text{spin}}(t).
\label{SpinNumbeStepB}
\end{equation}%
At the minimum we obtain 
\begin{equation}
\frac{\partial E_{\text{C}}}{\partial t}+\Delta _{\text{Z}}\frac{\partial N_{%
\text{spin}}}{\partial t}=0,  \label{SpinNumbeStepA}
\end{equation}%
from which we solve out $t=t_{0}(\Delta _{\text{Z}})$ and substituting it
back into (\ref{SpinNumbeStepB}), 
\begin{subequations}
\label{SpinNumbeStepD}
\begin{align}
E_{\text{sky}}(\Delta _{\text{Z}})=& E_{\text{C}}\left[ t_{0}(\Delta _{\text{%
Z}})\right] +\Delta _{\text{Z}}N_{\text{spin}}\left[ t_{0}(\Delta _{\text{Z}%
})\right] , \\
N_{\text{spin}}(\Delta _{\text{Z}})=& N\left[ t_{0}(\Delta _{\text{Z}})%
\right] .
\end{align}%
Using (\ref{SpinNumbeStepA}) it is easy to verify that 
\end{subequations}
\begin{equation}
N_{\text{spin}}(\Delta _{\text{Z}})=\frac{dE_{\text{sky}}(\Delta _{\text{Z}})%
}{d\Delta _{\text{Z}}},
\end{equation}%
which is (\ref{SpinNumbeEnerg}).

\begin{figure}[h]
\includegraphics[width=0.45\textwidth]{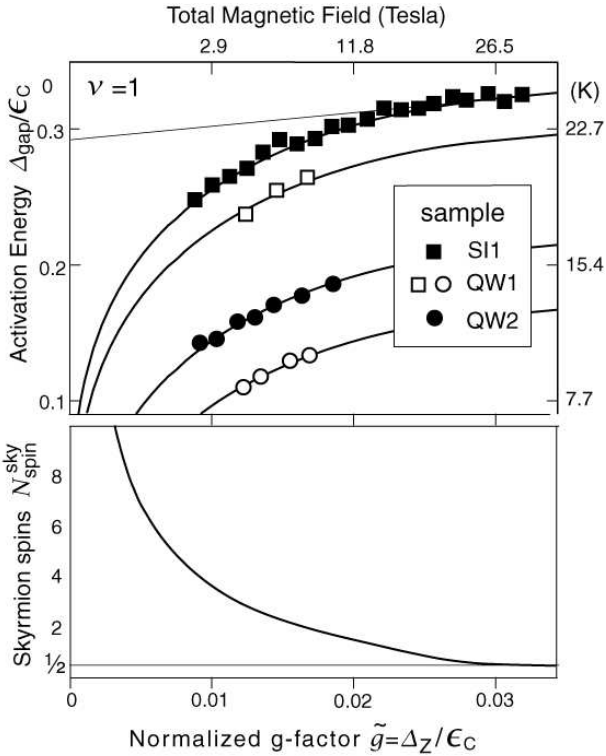}
\caption{A theoretical result on the activation energy $\Delta _{\text{gap}}$%
\ of a skyrmion-antiskyrmion pair is compared with experimental data at $%
\protect\nu =1$. The data are taken from Schmeller et. al..\protect\cite%
{Schmeller95L} The theoretical curve is based on the formula (\protect\ref%
{ActivEnergFormu}) with $\protect\gamma =0.56$. We have taken $\Gamma _{%
\text{offset}}=0.41$ for sample SI1. The offset $\Gamma _{\text{offset}}$
increases as the mobility decreases. The skyrmion spin is one half of the
slope of the activation-energy curve, $N_{\text{spin}}=\frac{1}{2}\partial
\Delta _{\text{gap}}/\partial \widetilde{g}$, where $\Delta _{\text{gap}}$
is taken in units of $\protect\epsilon _{\text{C}}$. The number $N_{\text{%
spin}}$ depends sensitively on the normalized Zeeman gap $\widetilde{g}$ for
small $\widetilde{g}$. The thin line is for the hole excitation energy. It
is seen that a hole is practically excited for $\widetilde{g}>0.03$.}
\label{FigSkyEne56}
\end{figure}

To compare our theoretical result with experimental data, it is necessary to
take into account two points so far neglected.

First, what we observe experimentally is the thermal activation energy of a
skyrmion-antiskyrmion pair. But this activation takes place in the presence
of charged impurities. The existence of charged impurities reduces the
activation energy considerably. We include an offset parameter $\Gamma _{%
\text{offset}}$ to treat this effect phenomenologically.\cite{Sasaki99B}

Second, we have so far assumed an ideal two-dimensional space for electrons.
This is not the case. Electrons are confined within a quantum well of a
finite width. This modifies the Coulomb energy considerably. The Coulomb
energy becomes smaller than what we have assumed. It is quite difficult to
make a rigorous analysis of the Coulomb energy in an actual quantum well. We
simulate the effect by including the reduction factor $\gamma $.

We consider the excitation energy of a skyrmion-antiskyrmion pair since it
is an observable quantity. It is simply twice of the skyrmion excitation
energy. Taking into account these two points, instead of (\ref%
{SpinNumbeStepB}) we set the activation energy as 
\begin{equation}
\Delta _{\text{gap}}(t)=2\gamma E_{\text{C}}(t)+2\Delta _{\text{Z}%
}N(t)-\Gamma _{\text{offset}},  \label{ActivEnergFormu}
\end{equation}%
where $0<\gamma <1$. Repeating the same steps as for (\ref{SpinNumbeStepD})
we come to 
\begin{subequations}
\label{SpinNumbeStepE}
\begin{align}
E_{\gamma }(\Delta _{\text{Z}})=& \gamma E_{\text{sky}}\left( \frac{\Delta _{%
\text{Z}}}{\gamma }\right) -\Gamma _{\text{offset}}, \\
N_{\gamma }(\Delta _{\text{Z}})=& N_{\text{spin}}\left( \frac{\Delta _{\text{%
Z}}}{\gamma }\right) =\frac{\partial E_{\gamma }(\Delta _{\text{Z}})}{%
\partial \Delta _{\text{Z}}},
\end{align}%
with the use of $E_{\text{sky}}(\Delta _{\text{Z}})$ and $N_{\text{spin}%
}(\Delta _{\text{Z}})$ derived in (\ref{SpinNumbeStepD}).

Experimental data\cite{Schmeller95L} were obtained in three samples (Fig.\ref%
{FigSkyEne56}): a single heterointerface (SI1) and two GaSa single quantum
wells (QW1 and QW2) with widths of 20nm and 14nm. The sample SI1 has a much
wider thickness of the two-dimensional sheet. In Fig.\ref{FigSkyEne56}, from
top to bottom the sample mobilities are $3.4$, $0.52$, $0.18$ and $%
0.16\times 10^{6}$ cm$^{-2}$, respectively. We have compared our theoretical
result (\ref{ActivEnergFormu}) successfully with the experimental data by
making appropriate choices of the reduction factor $\gamma $ and the offset
parameter $\Gamma _{\text{offset}}$ in Fig.\ref{FigSkyEne56}. We have used $%
\gamma =0.6$ for all samples. It would imply the thickness independence of
the excitation energy, suggesting that a skyrmion has a certain fixed size
in the third direction. However this problem is yet to be explored. On the
other hand we have used different values of $\Gamma _{\text{offset}}$ for
each. We have set $\Gamma _{\text{offset}}=0.4$ for sample SI1, and have
found that $\Gamma _{\text{offset}}$ increases as the mobility decreases.
This is reasonable since both the mobility and the offset $\Gamma _{\text{%
offset}}$ represent impurity effects.

\section{Semiclassical Approximation}

The formulas we have used to estimate the skyrmion excitation energy are
quite complicated. The modulations of the electron density and the spin
density, being expressed only in infinite power series as in (\ref%
{DensiCoulo}), are very difficult to handle with. Skyrmion excitations have
been observed also in bilayer QH systems\cite%
{Murphy94L,SawadaX03PE,Terasawa04}, where it is practically impossible to
repeat the calculations carried out here because too many dynamical
variables are involved. We have already applied a semiclassical
approximation to those systems\cite{EzawaX04B}. In this section we compare
the semiclassical result with the microscopic result.

In the semiclassical approximation we can make the spin-charge separation of
the electron field based on the composite-boson picture,\cite{Ezawa99L} 
\end{subequations}
\begin{equation}
\psi _{\mu }(\mathbf{x})=\psi (\mathbf{x})n_{\mu }(\mathbf{x}),
\end{equation}%
where the U(1) filed $\psi (\mathbf{x})$ carries the electric charge while
the CP$^{1}$ field $n_{\mu }(\mathbf{x})$ carries the spin. It leads to the
factorization of the spin field $S_{a}(\mathbf{x})$ into the number density $%
\rho (\mathbf{x})$ and the normalized spin field $\mathit{S}_{a}(\mathbf{x})$%
,%
\begin{equation}
S_{a}(\mathbf{x})=\rho (\mathbf{x})\mathit{S}_{a}(\mathbf{x})
\end{equation}%
with%
\begin{equation}
\mathit{S}_{a}(\mathbf{x})=\frac{1}{2}n_{\mu }^{\dag }(\mathbf{x})\left(
\tau _{a}\right) _{\mu \nu }n_{\nu }(\mathbf{x}).
\end{equation}%
The CP$^{1}$ field of the skyrmion is given by\cite{Ezawa99L}%
\begin{equation}
n_{\nu }(\mathbf{x})=\frac{1}{\sqrt{\left\vert z\right\vert ^{2}+\lambda ^{2}%
}}\left( 
\begin{array}{c}
z \\ 
\lambda%
\end{array}%
\right)
\end{equation}%
together with the normalized spin field (\ref{StandSkyrm}), and the wave
function is factorizable as in (\ref{FactoSkyrmWave}). We may also derive
the soliton equation\cite{Ezawa99L}%
\begin{equation}
{\frac{1}{4\pi }}\mathbf{\nabla }^{2}\ln \rho ^{\text{cl}}(\mathbf{x})-\rho
^{\text{cl}}(\mathbf{x})+\rho _{0}=\frac{2\lambda ^{2}\ell _{B}^{2}}{\left(
r^{2}+\lambda ^{2}\right) ^{2}}\rho _{0},  \label{SolitEq}
\end{equation}%
which determines the density modulation around the charge excitation.

We evaluate the excitation energy in the lowest order of $\ell _{B}^{2}$ by
assuming a large-scale skyrmion. It corresponds to the commutative limit in
the microscopic scheme.

The Coulomb energy is given by (\ref{DecomDX}). To calculate the direct
energy (\ref{DecomD}) we take the leading term in the solution of the
soliton equation (\ref{SolitEq}),%
\begin{equation}
\delta \rho ^{\text{cl}}(\mathbf{x})\simeq \frac{2\lambda ^{2}\ell _{B}^{2}}{%
\left( r^{2}+\lambda ^{2}\right) ^{2}}\rho _{0},
\end{equation}%
which agrees with the leading term in the density modulation (\ref%
{SkyrmStepEa}) in the microscopic theory. The direct Coulomb energy (\ref%
{DecomD}) reads%
\begin{equation}
H_{\text{D}}^{\text{cl}}=\pi \int \!d^{2}k\,V(\mathbf{k})e^{-\ell _{B}^{2}%
\mathbf{k}^{2}/2}\delta \rho ^{\text{cl}}(-\mathbf{k})\delta \rho ^{\text{cl}%
}(\mathbf{k})={\frac{3\pi ^{2}}{128\kappa }}\epsilon _{\text{C}},
\label{SkyrmEnergD}
\end{equation}%
where $\kappa =\lambda /2\ell _{B}$.

In calculating the exchange energy (\ref{DecomX}), since the exchange
potential $V_{\text{X}}(\mathbf{x})$ is short-ranged, we make the derivative
expansion and take the lowest order term. It is to make an approximation in (%
\ref{DecomX}),%
\begin{equation}
\delta V_{\text{X}}(\mathbf{k})\simeq -\frac{2J_{s}}{\pi \rho _{0}^{2}}%
\mathbf{k}^{2}
\end{equation}%
with (\ref{SpinStiff}). Furthermore we approximate $\rho ^{\text{cl}}(%
\mathbf{x})\simeq \rho _{0}$ in the exchange energy in the lowest order of $%
\ell _{B}^{2}$. Then the exchange energy turns out to be the nonlinear sigma
model,%
\begin{equation}
H_{\text{X}}^{\text{cl}}=\frac{J_{s}}{2}\sum_{a}\int \!d^{2}x\;[\partial _{k}%
\mathit{S}_{a}^{\text{cl}}(\mathbf{x})]^{2}.
\end{equation}%
This is bounded from below,%
\begin{equation}
H_{\text{X}}^{\text{cl}}\geq \pm 8\pi J_{s}Q_{\text{sky}}.
\end{equation}%
The lower bound, $H_{\text{X}}^{\text{cl}}=8\pi J_{s}$, is saturated by the
skyrmion configuration (\ref{StandSkyrm}). We note that the Hamiltonian $H_{%
\text{X}}^{\text{cl}}$ describes a spin wave. The coherence length $\xi $ is
given by%
\begin{equation}
\xi ^{2}=\frac{2J_{s}}{\rho _{0}\Delta _{\text{Z}}}=\frac{\sqrt{2\pi }}{8%
\widetilde{g}}\ell _{B}^{2}
\end{equation}%
in the presence of the Zeeman effect.

We cut off the divergence of the Zeeman energy. First, the skyrmion
excitation occurs within the coherent domain with the coherence length $\xi $%
. Second, the skyrmion size is proportional to the scale parameter $\kappa $%
. The Zeeman energy is approximated as 
\begin{equation}
\langle H_{\text{Z}}\rangle _{\text{sky}}=\frac{\lambda ^{2}\widetilde{g}}{%
2\ell _{B}^{2}}\ln \left( \frac{\alpha ^{2}\xi ^{2}}{4\ell _{B}^{2}}%
+1\right) ,
\end{equation}%
where we have cut off the upper limit of the integration at $r=\alpha \kappa
\xi $ with a phenomenological parameter $\alpha $.

Substituting these into (\ref{SkyrmEnergD}), we obtain%
\begin{equation}
\langle H\rangle _{\text{sky}}=\left[ {\frac{1}{4}}\sqrt{{\frac{\pi }{2}}}%
\beta +{\frac{3\pi ^{2}}{128\kappa }}+2\kappa ^{2}\widetilde{g}\ln \left(
\alpha ^{2}\frac{\sqrt{2\pi }}{32\widetilde{g}}+1\right) \right] \epsilon _{%
\text{C}},
\end{equation}%
where we have introduced another phenomenological parameter $\beta $ to
reduce the exchange energy, since the true exchange energy is smaller than $%
8\pi J_{s}$. Minimizing this with respect to $\kappa $ analytically we
obtain 
\begin{equation}
\langle H\rangle _{\text{sky}}\simeq \left\{ {\frac{1}{4}}\sqrt{{\frac{\pi }{%
2}}}\beta +{\frac{9\pi ^{2}}{256\kappa }}\right\} \epsilon _{\text{C}},
\label{SimplResul}
\end{equation}%
with%
\begin{equation}
\kappa \simeq {\frac{1}{2}}\left( {\frac{3\pi ^{2}}{64}}\right)
^{1/3}\left\{ \widetilde{g}\ln \left( \alpha ^{2}{\frac{\sqrt{2\pi }}{32%
\widetilde{g}}}+1\right) \right\} ^{-1/3}.  \label{OptimScale}
\end{equation}%
We have plotted the result in Fig.\ref{FigSkyEneSpinTh}. It reproduces
excellently the previous numerical result for $\widetilde{g}<0.038$ with the
choice of $\alpha =1.4$ and $\beta =0.9$. We may practically use it even in
the region where a hole is excited. We have thus confirmed the validity of
the semiclassical approximation.

\section{Conclusion}

\label{SecDiscu}

In this paper we have presented a microscopic theory of skyrmions in QH
ferromagnets. We have shown that the skyrmion is a W$_{\infty }$(2)-rotated
state of a hole-excited state. Because of an intrinsic entanglement between
the electron density and the spin density implied by the W$_{\infty }$(2)
algebra, a W$_{\infty }$(2) rotation modulates not only the spin
configuration but also the electron density around a hole, thus decreasing
the Coulomb energy. Similarly, the antiskyrmion is a W$_{\infty }$%
(2)-rotated state of an electron-excited state. There is a simple type of
skyrmion state characterized by the fact that its wave function is
factorizable in the electron coordinates. We call it the factorizable
skyrmion. It corresponds to the nonlinear-sigma-model skyrmion previously
derived in the semiclassical approximation. We have analyzed the skyrmion
state in the realistic Coulomb system with the Zeeman interaction. By
minimizing the excitation energy we have estimated the activation energy of
a skyrmion-antiskyrmion pair. The result is found to explain the
experimental data\cite{Schmeller95L} remarkably well.

\section{Acknowledgments}

We are grateful to the hospitality of Theoretical Physics Laboratory, RIKEN,
where a part of this work was done. On of the authors (ZFE) is supported in
part by Grants-in-Aid for Scientific Research from Ministry of Education,
Science, Sports and Culture (Nos. 13135202,14540237). The other author (GT)
acknowledges a research fellowship from Japan Society for Promotion of
Science (Nos. L04514).

\appendix

\section{Spontaneous Symmetry Breaking}

\label{AppSpontSymme}

We analyze the problem of spontaneous symmetry breaking due to a repulsive
interaction between electrons in the $\nu =1$ QH system. We assume that
every Landau site is occupied by one electron. The problem is to show that
the spin polarized state is the lowest-energy state though the Hamiltonian (%
\ref{SpinCoulo}) involves no spin variables.

We start with a proof, which makes the physical mechanism of spontaneous
symmetry breaking clear. According to the decomposition formula (\ref%
{DecomFormuA}) the energy of the four-fermion interaction Hamiltonian (\ref%
{SpinCoulo}) is a sum of the direct and exchange energies (\ref{DecomDX}),
or $\langle \mathfrak{S}|H_{\text{V}}|\mathfrak{S}\rangle =H_{\text{D}}^{%
\text{cl}}+H_{\text{X}}^{\text{cl}}$ with 
\begin{subequations}
\label{DecomDXx}
\begin{align}
H_{\text{D}}^{\text{cl}}=& \pi \int \!d^{2}k\,V(\mathbf{k})e^{-\frac{1}{2}%
\ell _{B}^{2}k^{2}}\left\vert \delta \hat{\rho}^{\text{cl}}(\mathbf{k}%
)\right\vert ^{2},  \label{AppenSSBa} \\
H_{\text{X}}^{\text{cl}}=& \pi \int \!d^{2}k\,\delta V_{\text{X}}(\mathbf{k})%
\left[ \left\vert \hat{S}_{a}^{\text{cl}}(\mathbf{k})\right\vert ^{2}+\frac{1%
}{4}\left\vert \hat{\rho}^{\text{cl}}(\mathbf{k})\right\vert ^{2}\right] ,
\end{align}%
where $V(\mathbf{k})>0$ and $\delta V_{\text{X}}(\mathbf{k})>0$ for a
repulsive interaction. Here, $\hat{\rho}^{\text{cl}}(\mathbf{k})=\langle 
\mathfrak{S}|\hat{\rho}(\mathbf{k})|\mathfrak{S}\rangle $ and $\hat{S}_{a}^{%
\text{cl}}(\mathbf{k})=\langle \mathfrak{S}|\hat{S}_{a}(\mathbf{k})|%
\mathfrak{S}\rangle $. The key observation is that, though the Hamiltonian (%
\ref{SpinCoulo}) involves no spin variables, the energy of a state does. It
is important that both the energies are positive semidefinite. The direct
energy $H_{\text{D}}^{\text{cl}}$ is insensitive to spin orientations, and
it vanishes for the homogeneous electron distribution since $\delta \hat{\rho%
}^{\text{cl}}(\mathbf{k})=0$. The exchange energy $H_{\text{X}}^{\text{cl}}$
depends on spin orientations. The spin texture is homogeneous when the spin
is completely polarized, where $\hat{S}_{a}^{^{\text{cl}}}(\mathbf{k}%
)\varpropto \delta (\mathbf{k})$. Furthermore, $\hat{\rho}^{\text{cl}}(%
\mathbf{k})\varpropto \delta (\mathbf{k})$ due to the homogeneous electron
distribution. For such a spin orientation the exchange energy also vanishes
since $\delta V_{\text{X}}(\mathbf{k})=V_{\text{X}}(\mathbf{k})-V_{\text{X}%
}(0)=0$ in (\ref{AppenSSBa}). On the other hand, $H_{\text{X}}^{\text{cl}}>0$
if the spin is not polarized completely since $\hat{S}_{a}^{^{\text{cl}}}(%
\mathbf{k})$ contains nonzero momentum components. Consequently the
spin-polarized state has the lowest energy, which is zero. Hence the
exchange interaction is the driving force of spontaneous symmetry breaking.

We may present another proof, which is mathematically more formal. First of
all, the state 
\end{subequations}
\begin{equation}
|\text{g}\rangle =\prod\limits_{n}c_{\uparrow }^{\dag }(n)|0\rangle
\end{equation}%
is an eigenstate of the Hamiltonian (\ref{SpinCoulo}) with the zero energy, $%
H_{\text{V}}|$g$\rangle =0$. It is one of the ground states. We shall prove
the following: (a) any globally W$_{\infty }$(2)-rotated state is degenerate
with $|$g$\rangle $; (b) any locally W$_{\infty }$(2)-rotated state has a
positive energy.

A spin rotated state of $|$g$\rangle $ is $|\Omega \rangle =e^{-i\Omega }|$g$%
\rangle $ with%
\begin{equation}
\Omega =\sum_{a}\sum_{mn}\xi _{a}(m,n)S_{a}(n,m),
\end{equation}%
where $\xi _{a}(m,n)$ is a Hermitian matrix, $[\xi _{a}(m,n)]^{\ast }=\xi
_{a}(n,m)$. We calculate how the rotation $e^{i\Omega }$ affects the
polarized spin for an infinitesimal parameter $\xi _{a}$. Using%
\begin{equation}
2\langle \text{g}|S_{a}(m,n)|\text{g}\rangle =\delta _{az}\delta _{mn},
\end{equation}%
we get 
\begin{equation}
2\langle \Omega |S_{a}(m,n)|\Omega \rangle =\delta _{az}\delta
_{mn}+\sum_{bc}\epsilon _{abc}\delta _{cz}\xi _{b}(m,n).
\end{equation}%
Only relevant transformations are generated by $\xi _{x}(m,n)$ and $\xi
_{y}(m,n)$, since $\xi _{z}(m,n)$ does not affect the spin polarization.
Besides we are interested in transformations rotating spins without moving
electrons from site to site. Then the parameter is reduced to $\xi
_{a}(m,n)=\xi _{a}(m)\delta _{mn}$.

The energy induced by an infinitesimal transformation is 
\begin{equation}
2\langle \Omega |H_{\text{V}}|\Omega \rangle =\sum_{a=x,y}\sum_{mn}\left[
\epsilon _{\text{X}}\delta _{mn}-V_{mnnm}\right] \xi _{a}(m)\xi _{a}(n),
\end{equation}%
where we have set $\langle $g$|H_{\text{V}}|$g$\rangle =0$. The question is
the positive definiteness of the quantity 
\begin{equation}
E(\xi )=\left[ \varepsilon _{\text{X}}\delta _{kn}-V_{kn}\right] \xi _{k}\xi
_{n},  \label{AppStepSponA}
\end{equation}%
where $V_{kn}\equiv V_{knnk}$ is the symmetric matrix.

Without loss of generality we assume the site index to run up to a finite
value $N_{\Phi }$, which will eventually be taken to infinity. The analysis
of (\ref{AppStepSponA}) is reduced to the analysis of the eigenvalues of $%
V_{kn}$. Introducing the complex valued function 
\begin{equation}
\varphi (\mathbf{x},\mathbf{y};\xi )=\sum_{k}\xi _{k}\varphi _{k}^{\ast }(%
\mathbf{x})\varphi _{k}(\mathbf{y}),
\end{equation}%
we write 
\begin{equation}
\sum_{kn}V_{kn}\xi _{k}\xi _{n}=\frac{1}{2}\int d^{2}xd^{2}yV(\mathbf{x}-%
\mathbf{y})\left\vert \varphi (\mathbf{x},\mathbf{y};\xi )\right\vert ^{2}>0.
\end{equation}%
Hence, $V_{kn}$ is positive definite, and all of its eigenvalues are
positive.

Now we refer to Gerschgorin's theorem\cite{BookGradshteyn}, which originally
deals with complex matrices. Formulating for a real matrix with positive
elements $a_{kn}>0$ the theorem reads as follows. Let $\Lambda _{k}$ be the
numbers defined as 
\begin{equation}
\Lambda _{k}=\sum_{n\neq k}a_{kn}>0.
\end{equation}%
Then all of the eigenvalues of $a_{kn}$ lie in the union of the segments $%
[a_{kk}-\Lambda _{k}\hspace{0.25mm},\hspace{0.25mm}a_{kk}+\Lambda _{k}]$.

In our case we have 
\begin{equation}
\Lambda _{k}=\sum_{n\neq k}V_{kn}=\sum_{n}V_{kn}-V_{kk}=\varepsilon _{\text{X%
}}-V_{kk}
\end{equation}%
so that the segments appear as $[2V_{kk}-\varepsilon _{\text{X}},\varepsilon
_{\text{X}}]$. On the other hand, we have shown that the eigenvalues of $%
V_{kn}$ are positive. Consequently, irrespective of the value of $%
2V_{kk}-\varepsilon _{\text{X}}$, we conclude that they lie within the
segment $[0,\varepsilon _{\text{X}}]$, meaning that the eigenvalues $E_{k}$
of the quadratic form (\ref{AppStepSponA}) satisfy the condition 
\begin{equation}
0\leqslant E_{k}\leqslant \varepsilon _{\text{X}}.
\end{equation}%
The eigenstate of the lowest eigenvalue $E=0$ is given by $\xi _{k}=\xi $,%
\begin{equation}
\sum_{n}\left[ \varepsilon _{X}\delta _{kn}-V_{kn}\right] \xi _{n}=\xi
\sum_{n}\left[ \varepsilon _{X}\delta _{kn}-V_{kn}\right] =0,
\end{equation}%
which corresponds to a global rotation of electron spins, while for an
arbitrary local rotation we have $E(\xi )>0$. It implies the degeneracy of
the ground states only under a global rotation, leading to a spontaneous
symmetry breaking of the rotational symmetry.

\section{Antiskyrmions}

\label{SecAntiSkyrm}

An antiskyrmion is a W$_{\infty }$($2$)-rotated state of the
electron-excited state. We consider the antiskyrmion state (\ref%
{MicroSkyrmStateAnti}), 
\begin{align}
|\mathfrak{S}_{\text{sky}}^{+}\rangle =& c_{\downarrow ,0}^{\dag }\left[
u_{0}^{+}c_{\uparrow ,0}^{\dag }+v_{0}^{+}c_{\downarrow ,1}^{\dag }\right] %
\left[ u_{1}^{+}c_{\uparrow ,1}^{\dag }+v_{1}^{+}c_{\downarrow ,2}^{\dag }%
\right]  \notag \\
\cdots & \left[ u_{N-1}^{+}c_{\uparrow ,N-1}^{\dag
}+v_{N-1}^{+}c_{\downarrow ,N}^{\dag }\right] c_{\uparrow ,N}^{\dag
}|0\rangle  \label{AntiStepA}
\end{align}%
with $(u_{n}^{+})^{2}+(v_{n}^{+})^{2}=1$. We note that the state comprises $%
N+2$ electrons over $N+1$ sites. Here and hereafter, for notational
simplicity, we set $c_{\mu n}\equiv c_{\mu }(n)$, $u_{n}^{+}\equiv u^{+}(n)$
and so on.

In contrast with the skyrmion state (\ref{MicroSkyrmState}), the
antiskyrmion state $|\mathfrak{S}_{\text{sky}}^{+}\rangle $ does not lead to
a wave function with any reasonable structure like (\ref{SkyrmWaveFunct}).
The expression for general $N$ is technically difficult to write down. As an
example we present the one for $N=2$. Up to multiplicative factors it
appears as 
\begin{subequations}
\begin{align}
\mathfrak{S}_{\uparrow \uparrow \uparrow \uparrow }^{+}=& 0,
\label{AntiStep1} \\
\mathfrak{S}_{\downarrow \uparrow \uparrow \uparrow }^{+}=&
(z_{2}-z_{3})(z_{2}-z_{4})(z_{3}-z_{4}), \\
\mathfrak{S}_{\downarrow \downarrow \uparrow \uparrow }^{+}=&
(z_{1}-z_{2})(z_{3}-z_{4})  \notag \\
& \times \left[ z_{3}z_{4}-(z_{1}+z_{2})(z_{3}+z_{4})\right] , \\
\mathfrak{S}_{\downarrow \downarrow \downarrow \uparrow }^{+}=&
(z_{1}-z_{2})(z_{1}-z_{3})(z_{2}-z_{3})z_{4}^{2}, \\
\mathfrak{S}_{\downarrow \downarrow \downarrow \downarrow }^{+}=& 0,
\label{AntiStep6}
\end{align}%
where no reasonable order is seen.

In this respect skyrmions and antiskyrmions might be regarded to possess
rather different properties. Actually, this is not so since the above scheme
of dealing with antiskyrmions is not quite satisfactory because of the
following observation.

The skyrmion state (\ref{MicroSkyrmState}) comprises $N$ electrons, and the
wave function (\ref{SkyrmWaveFunct}) describes these electrons in terms of
its $N$ complex arguments. On the other hand, the antiskyrmion wave function
comprises $N+2$ complex arguments, which at first sight seems to match the
number of electrons in the corresponding Fock state. However, since two
electrons out of those $N+2$ are fixed at $n=0$ and $n=N$, they cannot be
transferred to the neighboring sites. Therefore, though constructed of $N+2$
electrons, the antiskyrmion configuration comprises only $N-1$
site-transferrable electrons, which is exactly the same as in the skyrmion
state. In other word, the skyrmion and antiskyrmion Fock states comprise the
equal number of the degrees of freedom, since the two fixed electrons carry
no degrees of freedom at all. Nevertheless, the antiskyrmion wave function
involves $N+2$ complex arguments, which does not match the number of
site-transferable electrons. There is nothing wrong there, and the answer to
this mismatch is that the components of antiskyrmion wave function are not
independent, but subject to certain functional relations. The most
transparent manifestation of this statement are the relations (\ref%
{AntiStep1}) and (\ref{AntiStep6}).

Though the scheme is not wrong by itself, it is inconvenient to work with.
All these difficulties disappear if we develop a dual picture based on the
electron-hole symmetry. So far the antiskyrmion Fock state is built up via
creating electrons in the vacuum state. The dual picture deals with the same
state via removing electrons from the completely filled system of $N+1$
sites. Namely, up to the overall sign, the state (\ref{AntiStepA}) can be
rewritten as 
\end{subequations}
\begin{align}
|\mathfrak{S}_{\text{sky}}^{+}\rangle =& \left[ v_{0}^{+}c_{\uparrow
,0}-u_{0}^{+}c_{\downarrow ,1}\right] \left[ v_{1}^{+}c_{\uparrow
,1}-u_{1}^{+}c_{\downarrow ,2}\right]  \notag \\
& \cdots \left[ v_{N-1}^{+}c_{\uparrow ,N-1}-u_{N-1}^{+}c_{\downarrow ,N}%
\right] \prod_{\mu n}c_{\mu n}^{\dagger }|0\rangle .
\end{align}%
We now introduce the hole annihilation operators $\tilde{c}_{\mu n}$ and the
hole vacuum $|\tilde{0}\rangle $, 
\begin{equation*}
\tilde{c}_{\uparrow \downarrow ,n}=c_{\downarrow \uparrow ,n}^{\dag }%
\hspace*{5mm}\hspace*{5mm}|\tilde{0}\rangle =\prod_{\mu n}c_{\mu n}^{\dag
}|0\rangle
\end{equation*}%
with $\tilde{c}_{\mu n}|\tilde{0}\rangle =0$ and $\{\tilde{c}_{\mu m},\tilde{%
c}_{\nu n}^{\dag }\}=\delta _{\mu \nu }\delta _{mn}$.

Now the antiskyrmion Fock state can be presented as%
\begin{align}
|\mathfrak{S}_{\text{sky}}^{+}\rangle =& \left[ v_{0}^{+}\tilde{c}%
_{\downarrow ,0}^{\dagger }-u_{0}^{+}\tilde{c}_{\uparrow ,1}^{\dagger }%
\right] \left[ v_{1}^{+}\tilde{c}_{\downarrow ,1}^{\dagger }-u_{1}^{+}\tilde{%
c}_{\uparrow ,2}^{\dagger }\right]  \notag \\
& \cdots \left[ v_{N-1}^{+}\tilde{c}_{\downarrow ,N-1}^{\dagger }-u_{N-1}^{+}%
\tilde{c}_{\uparrow ,\hspace*{0.25mm}N}^{\dagger }\right] |\tilde{0}\rangle ,
\end{align}%
where the analogy with the skyrmion state is manifest.

In the same way we introduce the hole field operator as $\tilde{\psi}%
_{\uparrow \downarrow }(\mathbf{r})=\psi _{\downarrow \uparrow }^{\dag }(%
\mathbf{r})$, or equivalently%
\begin{equation}
\tilde{\psi}_{\mu }(\mathbf{x})=\rho _{0}^{1/2}e^{-|\bar{z}%
|^{2}/4}\sum_{n=0}\alpha (n)\bar{z}^{n}\tilde{c}_{\mu }(n)
\end{equation}
where $\bar{z}=\left( x-iy\right) /\ell _{B}$.

Now the wave function of the antiskyrmion state comprising $N+2$ electrons
contains $N$ variables $\bar{z}_{1},\ldots ,\bar{z}_{N}$. So the unphysical
degrees of freedom associated with two non-transferable electrons of (\ref%
{AntiStepA}) do not appear at all, since those two electrons are accounted
within the ground state $|\tilde{0}\rangle $.

Using the analogy with (\ref{SkyrmWaveFunct}) it is straightforward to write
down the $N$-body wave function of antiskyrmion configuration. We obtain 
\begin{align}
& \mathfrak{S}_{\mu _{1}\mu _{2}\cdots \mu _{N}}^{+}[\mathbf{x}%
]=C_{N}e^{-\sum_{r=1}^{N-1}|z_{r}|^{2}/4}  \notag \\
& \times \left\vert 
\begin{array}{cccc}
\bar{z}_{1}^{0}\left( 
\begin{array}{c}
\bar{z}_{1} \\ 
\tilde{\kappa}_{0}%
\end{array}%
\right) _{\mu _{1}} & \bar{z}_{1}^{1}\left( 
\begin{array}{c}
\bar{z}_{1} \\ 
\tilde{\kappa}_{1}%
\end{array}%
\right) _{\mu _{1}} & \cdots & \bar{z}_{1}^{N-1}\left( 
\begin{array}{c}
\bar{z}_{1} \\ 
\tilde{\kappa}_{N-1}%
\end{array}%
\right) _{\mu _{1}} \\ 
\bar{z}_{2}^{0}\left( 
\begin{array}{c}
\bar{z}_{2} \\ 
\tilde{\kappa}_{0}%
\end{array}%
\right) _{\mu _{2}} & \bar{z}_{2}^{1}\left( 
\begin{array}{c}
\bar{z}_{2} \\ 
\tilde{\kappa}_{1}%
\end{array}%
\right) _{\mu _{2}} & \cdots & \bar{z}_{2}^{N-1}\left( 
\begin{array}{c}
\bar{z}_{2} \\ 
\tilde{\kappa}_{N-1}%
\end{array}%
\right) _{\mu _{2}} \\ 
\vdots & \hspace*{-1mm}\hspace*{-1mm}\vdots & \ddots & \vdots \\ 
\bar{z}_{N}^{0}\left( 
\begin{array}{c}
\bar{z}_{N} \\ 
\tilde{\kappa}_{0}%
\end{array}%
\right) _{\mu _{N}} & \bar{z}_{N}^{1}\left( 
\begin{array}{c}
\bar{z}_{N} \\ 
\tilde{\kappa}_{1}%
\end{array}%
\right) _{\mu _{N}} & \cdots & \bar{z}_{N}^{N-1}\left( 
\begin{array}{c}
\bar{z}_{N} \\ 
\tilde{\kappa}_{N-1}%
\end{array}%
\right) _{\mu _{N}}%
\end{array}%
\right\vert  \label{AntiStepB}
\end{align}%
as in (\ref{SkyrmWaveFunct}).

\section{Decomposition Formula}

\label{SecDecomFormu}

We prove the decomposition formula (\ref{DecomFormuA}), according to which
the four-fermion interaction energy consists of the direct energy and the
exchange energy. We evaluate the energy of a state $|\mathfrak{S}\rangle $
in the class of states (\ref{GenerState}),%
\begin{equation}
\langle H_{\text{V}}\rangle =\langle \mathfrak{S}|H_{\text{V}}|\mathfrak{S}%
\rangle ,
\end{equation}%
where the Hamiltonian is given by (\ref{SpinCoulo}). We assume that the
electron field carries the SU(N$_{\text{f}}$) isospin index, $\mu
=1,2,\ldots ,N_{\text{f}}$. For the sake of accuracy we first consider a
system with a finite number of Landau sites ($m=0,1,\ldots ,N_{\Phi }-1$)
and take the limit $N_{\Phi }\rightarrow \infty $ in final expressions. It
is convenient to combine the isospin and site indices into a multi-index $%
M\equiv (\mu ,m)$, where the multi-index runs over the values $M=1,2,\ldots
,N_{\text{f}}N_{\Phi }$.

The W$_{\infty }$($N_{\text{f}}$) algebra is identical to the algebra U$(N_{%
\text{f}}N_{\Phi })$ in the limit $N_{\Phi }\rightarrow \infty $, and the
transformation rules for the fermion operators appear as 
\begin{align}
e^{-iW}c_{M}e^{iW}=& (U)_{MM^{\prime }}c_{M^{\prime }},  \notag \\
e^{-iW}c_{M}^{\dag }e^{iW}=& c_{M^{\prime }}^{\dag }(U^{\dag })_{M^{\prime
}M},  \label{DFa}
\end{align}%
where $U$ is an $(N_{\text{f}}N_{\Phi })\times (N_{\text{f}}N_{\Phi })$
unitary matrix, $UU^{\dag }=U^{\dag }U=\mathbb{I}_{(N_{\text{f}}N_{\Phi
})\times (N_{\text{f}}N_{\Phi })}$. Here and hereafter the repeated index
implies the summation over it.

We first calculate the two-point averages by the state $|\mathfrak{S}\rangle
=e^{iW}|\mathfrak{S}_{0}\rangle $ in (\ref{GenerState}). Using (\ref{DFa})
we get 
\begin{equation}
\langle \mathfrak{S}|c_{M}^{\dag }c_{N}|\mathfrak{S}\rangle =(U^{\dag
})_{KM}(U)_{NL}\langle \mathfrak{S}_{0}|c_{K}^{\dag }c_{L}|\mathfrak{S}%
_{0}\rangle .
\end{equation}%
For the state $|\mathfrak{S}_{0}\rangle $ given by (\ref{PureState}) we have 
\begin{equation}
\langle \mathfrak{S}_{0}|c_{K}^{\dag }c_{L}|\mathfrak{S}_{0}\rangle =\nu
_{K}\delta _{KL},
\end{equation}%
which eventually leads to 
\begin{equation}
\langle \mathfrak{S}|c_{M}^{\dag }c_{N}|\mathfrak{S}\rangle =\nu
_{K}(U)_{NK}(U^{\dag })_{KM}.  \label{DFd}
\end{equation}%
Carrying out analogous manipulations in four-point averages we get 
\begin{align}
\langle \mathfrak{S}|c_{M}^{\dag }c_{S}^{\dag }c_{T}c_{N}|\mathfrak{S}%
\rangle =& (U^{\dag })_{KM}(U^{\dag })_{IS}(U)_{TJ}(U)_{NL}  \notag \\
& \times \langle \mathfrak{S}_{0}|c_{K}^{\dag }c_{I}^{\dag }c_{J}c_{L}|%
\mathfrak{S}_{0}\rangle .  \label{DFb}
\end{align}%
We can use 
\begin{equation}
\langle \mathfrak{S}_{0}|c_{K}^{\dag }c_{I}^{\dag }c_{J}c_{L}|\mathfrak{S}%
_{0}\rangle =\nu _{J}\nu _{L}(\delta _{IJ}\delta _{KL}-\delta _{IL}\delta
_{KJ})  \label{DFc}
\end{equation}%
for the state $|\mathfrak{S}_{0}\rangle $. Substituting (\ref{DFc}) into (%
\ref{DFb}) and accounting (\ref{DFd}) we summarize as 
\begin{align}
\langle \mathfrak{S}|c_{M}^{\dag }c_{S}^{\dag }c_{T}c_{N}|\mathfrak{S}%
\rangle =& \langle \mathfrak{S}|c_{M}^{\dag }c_{N}|\mathfrak{S}\rangle
\langle \mathfrak{S}|c_{S}^{\dag }c_{T}|\mathfrak{S}\rangle  \notag \\
& -\langle \mathfrak{S}|c_{M}^{\dag }c_{T}|\mathfrak{S}\rangle \langle 
\mathfrak{S}|c_{S}^{\dag }c_{N}|\mathfrak{S}\rangle ,
\end{align}%
where the direct and exchange terms are easily recognized. Here, we have 
\begin{equation}
\langle \mathfrak{S}|c_{\mu }^{\dag }(m)c_{\nu }(n)|\mathfrak{S}\rangle =%
\frac{\delta _{\nu \mu }}{N_{\text{f}}}\rho ^{\text{cl}}(n,m)+(\lambda
_{A})_{\nu \mu }S_{A}^{\text{cl}}(n,m),  \label{DFe}
\end{equation}%
in terms of decoupled spin and site indices, where $\rho ^{\text{cl}}(m,n)$
and $S_{A}^{\text{cl}}(m,n)$ are defined by (\ref{ClassDensi}) with $\lambda
_{A}$ the Gell-Mann matrix, $A=1,\cdots ,N_{\text{f}}^{2}-1$.

In this way we achieve at the decomposition formula, 
\begin{equation}
\langle H_{\text{V}}\rangle =H_{\text{D}}^{\text{cl}}+H_{\text{X}}^{\text{cl}%
},
\end{equation}%
where the direct and the exchange parts are given by 
\begin{align}
H_{\text{D}}^{\text{cl}}=& V_{mnij}\rho ^{\text{cl}}(n,m)\rho ^{\text{cl}%
}(j,i)-(N_{\Phi }+2\Delta N)\epsilon _{\text{D}}, \\
H_{\text{X}}^{\text{cl}}=& -2V_{mnij}S_{A}^{\text{cl}}(j,m)S_{A}^{\text{cl}%
}(n,i)-\frac{1}{N_{\text{f}}}V_{mnij}\rho ^{\text{cl}}(j,m)\rho ^{\text{cl}%
}(n,i)  \notag \\
& +(N_{\Phi }+\delta N)\epsilon _{\text{X}}.
\end{align}%
They read (\ref{DecomDXx}) in the momentum representation.

In deriving the formula for $H_{\text{X}}^{\text{cl}}$ we have accounted $V_{%
\text{X}}(0)=4\ell ^{2}\epsilon _{X}$ and also the relation%
\begin{align}
& \frac{N_{\Phi }+\delta N}{4\pi \ell _{B}^{2}}=\frac{1}{4\pi \ell _{B}^{2}}%
\int \!d^{2}x\,\hat{\rho}^{\text{cl}}(\mathbf{x})  \notag \\
=& \int \!d^{2}k\,[\hat{S}_{A}^{\text{cl}}(-\mathbf{k})\hat{S}_{A}^{\text{cl}%
}(\mathbf{k})+\frac{1}{2N_{\text{f}}}\hat{\rho}^{\text{cl}}(-\mathbf{k})\hat{%
\rho}^{\text{cl}}(\mathbf{k})].  \label{DFf}
\end{align}%
This relation is derived as follows. We deal with the quantity 
\begin{equation}
\langle \mathfrak{S}|c_{M}^{\dag }c_{K}|\mathfrak{S}\rangle \langle 
\mathfrak{S}|c_{K}^{\dag }c_{N}|\mathfrak{S}\rangle =\nu
_{K}(U)_{NK}(U^{\dag })_{KM},
\end{equation}%
where we have used (\ref{DFd}) and $\nu _{K}^{2}=\nu _{K}$ since $\nu _{K}=0$
or $1$. Comparing this with (\ref{DFd}) we conclude 
\begin{equation}
\langle \mathfrak{S}|c_{M}^{\dag }c_{K}|\mathfrak{S}\rangle \langle 
\mathfrak{S}|c_{K}^{\dag }c_{N}|\mathfrak{S}\rangle =\langle \mathfrak{S}%
|c_{M}^{\dag }c_{N}|\mathfrak{S}\rangle ,
\end{equation}%
which is%
\begin{equation}
\langle \mathfrak{S}|c_{\mu }^{\dag }(m)c_{\kappa }(k)|\mathfrak{S}\rangle
\langle \mathfrak{S}|c_{\kappa }^{\dag }(k)c_{\nu }(n)|\mathfrak{S}\rangle
=\langle \mathfrak{S}|c_{\mu }^{\dag }(m)c_{\nu }(n)|\mathfrak{S}\rangle .
\label{BasicFormuNCa}
\end{equation}%
Substituting (\ref{DFe}) into this we find%
\begin{equation}
\hat{S}_{A}^{\text{cl}}(n,k)\hat{S}_{A}^{\text{cl}}(k,m)+\frac{1}{4N_{\text{f%
}}}\hat{\rho}^{\text{cl}}(n,k)\hat{\rho}^{\text{cl}}(k,m)=\frac{1}{2}\hat{%
\rho}^{\text{cl}}(n,m).
\end{equation}%
This amounts to (\ref{DFf}) in the momentum space.

\end{document}